\documentclass[letter,12pt]{article}
\usepackage[top=1in,bottom=1in,left=1in,right=1in]{geometry}

\pdfoutput=1

\usepackage{amsmath,amssymb,amsfonts}
\usepackage{amsthm}         
\usepackage{latexsym}
\usepackage[scr=boondox]{mathalfa}
\usepackage{dsfont}
\usepackage{physics}
\usepackage{slashed}
\usepackage[normalem]{ulem}
\usepackage{simpler-wick}
\usepackage{cancel}
\usepackage{yhmath}
\usepackage{mathdots}
\usepackage{cite}
\usepackage{setspace}
\usepackage{pgfplots}
\pgfplotsset{compat=1.17}
\usepackage{tikz}
\usepackage{tikz-3dplot}
\usetikzlibrary{
  3d,
  calc,
  fadings,
  patterns,
  decorations.pathreplacing,
  shadows.blur,
  shapes
}

\usepackage{booktabs}
\usepackage{multirow}
\usepackage{tabularx}
\usepackage{array}

\usepackage{gensymb}
\usepackage{siunitx}
\usepackage{pifont}

\usepackage[utf8]{inputenc}
\usepackage{verbatim}

\usepackage{hyperref}

\usepackage{caption}

\numberwithin{equation}{section}

\newcommand{\be}{\begin{equation}}
\newcommand{\ee}{\end{equation}}
\newcommand{\bs}{\begin{subequations}}
\newcommand{\es}{\end{subequations}}

\newcommand{\pa}{\partial}

\newcommand{\beq}{\begin{eqnarray}}
\newcommand{\eeq}{\end{eqnarray}}

\def\e{{\epsilon}}

\def\s{{\sigma}}

\def\a{{\alpha}}
\def\b{{\beta}}
\def\g{{\gamma}}
\def\G{{\Gamma}}
\def\Th{{\Theta}}
\def\d{{\delta}}
\def\ka{{\kappa}}
\def\th{{\theta}}
\def\D{\Delta}
\def\O{\Omega}

\def\th{\theta}

\def\CB{{\mathcal B}}

\def\CH{{\mathcal H}}

\def\CM{{\mathcal M}}

\def\CO{{\mathcal O}}

\def\A{{\mathcal A}}

\def\SI{{\mathscr I}}

\newcommand{\dt}{{\text d}}

\def\+{{(+)}}
\def\-{{(-)}}
\def\0{{(0)}}
\def\1{{(1)}}
\def\2{{(2)}}
\def\3{{(3)}}

\def\p{{\partial}}

\def\bw{{\bar{w}}}

\def\soft{{\text{soft}}}
\def\tu{{U}}
\def\tr{{R}}

\def\tx{{X}}
\def\tz{{Z}}
\def\tbz{{{\bar{Z}}}}

\def\rd{\text{d}}
\def\bz{{\bar{z}}}

\def\AFG{{\text{\tiny AFG}}}

\begin{document}
\begin{titlepage}
\unitlength = 1mm
\hfill CALT-TH 2026-023
\ \\
\vskip 2cm
\begin{center}

\openup .5em

{\LARGE{Mapping the Infrared Phase Space \\ of Gravity to Finite Subregions}}

\vspace{0.8cm}
Luca Ciambelli,$^1$ Temple He,$^2$ Marc S. Klinger,$^2$ Kathryn M. Zurek$^2$

\vspace{1cm}

{\it  $^1$Perimeter Institute for Theoretical Physics \\ 31 Caroline Street North, Waterloo, Ontario, Canada N2L 2Y5} \\
{\it  $^2$Walter Burke Institute for Theoretical Physics \& Leinweber Forum for Theoretical Physics \\ California Institute of Technology, Pasadena, CA 91125 USA}\\

\vspace{0.8cm}

\begin{abstract}
We construct the phase space for an arbitrary cut of a null hypersurface in Minkowski spacetime and demonstrate that it is symplectomorphic to the infrared phase space of asymptotically flat gravity. Fluctuations of the cut are mapped to the leading soft graviton mode. Furthermore, the supertranslation Goldstone mode is mapped to the product of the size of the cut with its symplectic partner, the null time offset. 
 \end{abstract}

\vspace{1.0cm}
\end{center}
\end{titlepage}
\pagestyle{empty}
\pagestyle{plain}
\pagenumbering{arabic}

\tableofcontents

\newpage

\section{Introduction}

There has been much recent progress in understanding the phase space associated to area fluctuations of finite causal diamonds \cite{Verlinde:2019xfb, Verlinde:2019ade, Ciambelli:2025flo, Bub:2024nan, He:2025hag, Ciambelli:2025fbo,Klinger:2023tgi,Klinger:2025hjp, Freidel:2026hed, Kowalski-Glikman:2026bwg}. In particular, \cite{Bub:2024nan} considered a toy model where only spherically symmetric ($S$-wave) area fluctuations of a finite causal diamond were included. Importantly, the phase space associated to this toy model localizes to a codimension-2 surface that lives at the bifurcate horizon. Consequently, the physical degrees of freedom contained in the phase space are referred to as edge modes \cite{Donnelly:2016auv,Speranza:2017gxd,Geiller:2019bti,He:2024ddb, Ball:2024hqe,Ball:2024xhf,Ball:2024gti, He:2024skc, Araujo-Regado:2024dpr, Carrozza:2022xut,Ciambelli:2021nmv,Ciambelli:2022vot,Klinger:2023qna,Klinger:2026tws}. The resulting symplectic form is characterized by a single conjugate pair featuring the horizon area $A$ and its symplectic partner $\a$ describing offsets in null time. 

A similar situation exists in the leading infrared sector of the phase space associated to asymptotically flat gravity, where the relevant degrees of freedom are the soft graviton mode $N$ and its symplectic conjugate, the supertranslation Goldstone mode $C$ \cite{He:2014laa, He:2023qha, He:2024vlp, AndradeeSilva:2026mpa}. In this case, both $N$ and $C$ are localized to the codimension-2 celestial sphere at $\SI^+_-$, the past boundary of $\SI^+$.\footnote{Equivalently, one may view the celestial sphere as living at $\SI^-_+$, the future boundary of $\SI^-$, via the antipodal matching conditions \cite{Strominger:2013jfa}.} Although the celestial sphere is formally infinitely large, one can still study renormalized finite area fluctuations, which are sourced by the soft graviton mode $N$ \cite{Kapec:2016aqd}.

It is natural to ask whether there is a formal relationship between these two phase spaces. This question was explored in \cite{Ciambelli:2025fbo}, where it was shown that there is indeed a symplectomorphism, i.e., an invertible mapping which preserves Poisson brackets, between the two phase spaces. By geometrically matching the area fluctuations induced by $N$ in the asymptotically flat phase space to fluctuations of the horizon area $A$ in the causal diamond, one can determine the precise mapping between $N$ and $A$. Using the symplectic structure as a guide, it is then clear how to map between $C$ and $\a$ so as to preserve the Poisson brackets.\footnote{Strictly speaking, there is an ambiguity in mapping $C$ to $\a$. This was explained in \cite{Ciambelli:2025fbo}, where a particular, albeit natural, choice was made. } However, because the phase space of the causal diamond considered in \cite{Bub:2024nan} only involves spherically symmetric perturbations, the identification described above can only be made after angle-averaging the asymptotic modes $N$ and $C$. Thus, while the existence of such a relationship between the two phase spaces is formally interesting, the presence of spherical symmetry limits the scope of the result. Any realistic two-armed interferometer experiment, which naturally defines a finite causal diamond, is sensitive to area fluctuations that are not spherically symmetric \cite{Verlinde:2019xfb,Li:2022mvy,Vermeulen:2024vgl}.

\begin{figure}[t]
\centering
\begin{tikzpicture}[scale=0.6]
    \draw[->] (0,-3) -- (0,3) node[above] {$t$};
    \draw[->] (-1,-2) -- (5,-2) node[right] {$r$};

    \draw[thick, blue, domain=1:3, samples=100, smooth] 
    plot (-{\x} + 3, {sqrt(\x^2 - 1)-2} );	

    \draw[thick, red] (0,2) -- (4,-2); 
    \node[red] at (1.5,1.5) {$\CH^+$ };
    \node[red] at (2.8, 0.5) {\small $\tr=0$};
    \node[blue] at (0.5,-1.3) {\small $\tr=\tr_0$};
    \node at (4,-2) [circle, fill, cyan, inner sep=1.5pt] {};
    \node[below right, cyan] at (4,-2) {$\CB \; (\a(z,\bz) ,\e(z,\bz))$};

     \draw[thick] (0,4.5) -- (4,4.5);
    \draw[thick] (0,4.4) -- (0,4.6); 
    \draw[dashed] (0,4.4) -- (0,4);
    \draw[thick] (4,4.4) -- (4,4.6);   
    \draw[dashed] (4,4.4) -- (4,-2);
    \node at (2, 5) { $L(z,\bz)$ };

    \draw[thick, red] (12,5) -- (17,0); 
    \draw[thick,black] (17,0) -- (12,-5) -- (12,5);
    \node[red] at (15,3) { $\SI^+$ };
    \node at (15,-3) { $\SI^-$ };
    \node at (17,0) [circle, fill, inner sep=1.5pt] {};
    \node at (12,5) [circle, fill, inner sep=1.5pt] {};
    \node at (12,-5) [circle, fill, inner sep=1.5pt] {};
    \node[below right] at (17,0) {$i^0$};
    \node[above] at (12,5) {$i^+$};
    \node[below] at (12,-5) {$i^-$};

    \draw[dashed] (17,0) -- (18.5,4);
    \draw[dashed] (17,0) -- (18.5,2);
    \draw[thick, black] (18.5,2) -- (18.5,4);
    \node at (18.5,4) [circle, fill, cyan, inner sep=1.5pt] {};
    \node[above, cyan] at (18.5,4) {$\SI^+_-  \; (C(z,\bz), N(z,\bz))$}; 
 
  \def\A{0.1}       
  \def\lambda{1.0}   
  \def\k{360/\lambda} 
  \draw[domain=0:2, samples=400, smooth] (2,0)
    plot (\x + 13 ,{-1.5 +  \A * sin(\k * \x) });
  \draw[domain=0:2, samples=400, smooth] (2,0)
    plot (\x + 14 ,{-0.3 +  \A * sin(\k * \x) });
  \draw[domain=0:2, samples=400, smooth] (2,0)
    plot (\x + 12.5 ,{1.5 +  \A * sin(\k * \x) });
    
\end{tikzpicture}

\caption{We illustrate the two relevant geometric frameworks we are relating in this paper in four bulk spacetime dimensions. On the left, we have projected out the angular dependence in drawing the spacetime diagram for the future half of a causal diamond in a Minkowski background. The past boundary of $\CH^+$ is chosen to be $\CB$, a fixed cut given by $L(z,\bz)$, where $(z,\bz)$ label the angular dependence. The phase space lives at $\CB$ and consists of $\e(z,\bz)$, the dynamical part of $L(z,\bz)$ controlling length fluctuations, and its symplectic partner $\a(z,\bz)$, the null time offset. We have also illustrated a constant $\tr=\tr_0$ hypersurface, which is known as a stretched horizon. On the right, we have the Penrose diagram of an asymptotically flat spacetime, with the ripples indicating fluctuations. The leading infrared sector of the phase space lives at $\SI^+_-$ and consists of the soft graviton mode $N(z,\bz)$ and its symplectic partner $C(z,\bz)$, the supertranslation Goldstone mode}. \label{fig:two-setups}
\end{figure}
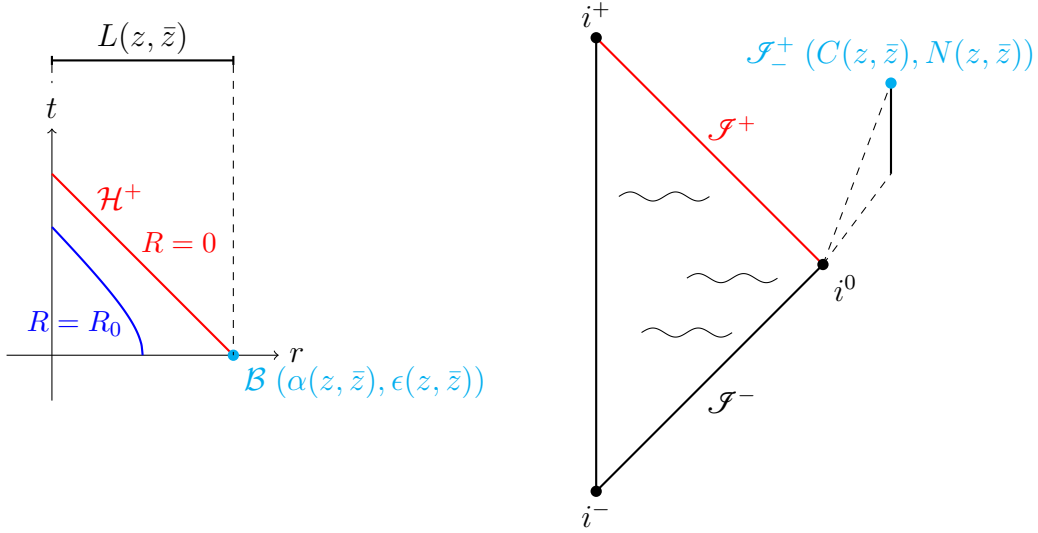

In this paper, we overcome this limitation by showing that the identification between the infrared phase space of asymptotically flat gravity and the phase space associated to a cut of the null horizon in a Minkowski background persists even after relaxing spherical symmetry. In this regard, we view this paper as a generalization of the analysis performed in \cite{Bub:2024nan, Ciambelli:2025fbo}. We have illustrated the two physical configurations in four bulk spacetime dimensions in Figure~\ref{fig:two-setups}. On the right is the Penrose diagram for an asymptotically flat spacetime (e.g., see \cite{Strominger:2017zoo}). We have blown up spatial infinity $i^0$ to illustrate that the physical modes comprising the infrared phase space, namely the soft graviton $N$ and the supertranslation Goldstone $C$, live on the codimension-2 celestial sphere located at $\SI^+_-$, the past boundary of $\SI^+$ (or equivalently the future boundary of $i^0$). The left diagram depicts the finite null future horizon $\CH^+$ of a lightcone, which we can take to be the future half of a causal diamond. It is ostensibly similar to its right counterpart, but there are a few subtleties. We choose the past boundary of $\CH^+$ to be an arbitrary cut, which we call $\CB$ (see Figure~\ref{fig:GNC} below). We characterize $\CB$ by an angle-dependent function $L(z,\bz)$, where $(z,\bz)$ are complex coordinates parametrizing the solid angle. If $\CB$ is the spherically symmetric cut located at $t=0$, so that $L(z,\bz) \equiv L$ is a constant, we recover the case considered in \cite{Bub:2024nan, Ciambelli:2025fbo}. The degrees of freedom controlling the fluctuations of the cut $\CB$ along $\CH^+$ are determined to be $\e(z,\bz)$, which is the dynamical part of $L(z,\bz)$, and its symplectic partner $\a(z,\bz)$, which is associated with offsets in null time. 

As we shall see, since both $N(z,\bz)$ and $\e(z,\bz)$ control area fluctuations in their respective phase spaces, we can make the identification (see \eqref{eps-N})
\begin{align}
	\frac{1}{4}\Box(\Box+2) N(z,\bz) \sim \e(z,\bz).
\end{align}
Following the argument given in \cite{Ciambelli:2025fbo}, we can then exploit the symplectic structure of the two phase spaces to further identify\footnote{In \cite{Bub:2024nan, Ciambelli:2025fbo}, the authors used the notation $\mu$ to denote the symplectic conjugate to $\e$. We refrain from using this notation since $\mu$ is conventionally used in the Carrollian physics community (e.g., see \cite{Hopfmuller:2018fni}) to denote the combination $\ka + \frac{d-1}{d}\th$, where $\ka$ is the inaffinity and $\th$ the expansion parameter.} (see \eqref{CL-fin})
\begin{align}
	C(z,\bz) \sim 2 \a(z,\bz) L(z,\bz).
\end{align}
These identifications generalize those found in \cite{Ciambelli:2025fbo} to include non-spherically symmetric fluctuations of the cut $\CB$, thereby allowing us to fully bridge the asymptotic analysis involving the soft modes with the edge modes corresponding to the past boundary of a finite null hypersurface in Minkowski spacetime. The length fluctuations $\e(z,\bz)$ are central to interferometer experiments~\cite{Verlinde:2022hhs,Zurek:2022xzl}, whereas the soft modes are responsible for the gravitational memory effect \cite{Strominger:2014pwa}. The identification we establish between soft and edge modes in this paper therefore suggests we may interpret such length fluctuations analogously as a gravitational memory effect. However, it is important to remark that some physical aspects differ in the two settings. Asymptotically, gravitational fluctuations do not backreact on the boundary geometry, while gravitational fluctuations of a finite-size causal diamond inevitably backreact and modify its geometry. For these reasons, we leave a precise connection between the memory effect and the length fluctuations for future work.

The paper is organized as follows. In Section~\ref{sec:phase-space-null}, we review some basic properties of null hypersurfaces and construct the on-shell phase space for the past boundary of a finite null hypersurface in Minkowski spacetime.  Next, in Section~\ref{sec:connection}, we first give a brief review of asymptotically flat spacetimes and their associated phase space in the leading soft limit. We then describe a symplectomorphism between this asymptotic phase space and the phase space constructed in Section~\ref{sec:phase-space-null}. Finally, we conclude in Section~\ref{sec:discussion} with a summary of our main results and some discussion of future directions.

\section{Phase space for finite null hypersurfaces}
\label{sec:phase-space-null}

In this section, we analyze the on-shell phase space for area fluctuations of an arbitrary cut of a null hypersurface in a Minkowski background. The Carrollian perspective is very useful to this end, and we provide in Appendix~\ref{app:carroll-review} a quick summary of the geometry of null hypersurfaces, which is based on the recent review \cite{Ciambelli:2025unn}. Practically, it suffices for our purposes to restrict ourselves to a special subclass of general Carrollian spacetimes which can be described using Gaussian null coordinates. In Section~\ref{subsec:carroll}, we will construct the on-shell phase space associated to a null hypersurface in an arbitrary background by imposing the Raychaudhuri and Damour equations. We will then specialize to the case of a Minkowski background in Section~\ref{subsec:mink}, where the simplicity of the spacetime allows us to give a geometric interpretation to the dynamical modes in phase space. This will be important when we utilize a geometric argument to connect this phase space to that of asymptotically flat spacetimes in Section~\ref{sec:connection}.

\begin{figure}[t]
\centering
\tdplotsetmaincoords{70}{110}
\begin{tikzpicture}[scale=4,tdplot_main_coords]
  \def\a{3}
  \def\b{1.5}

  \coordinate (Apex) at (0,0,1.25);

\shade[shading=ball, ball color=gray!65, opacity=0.85]
  plot[smooth cycle, samples=180, variable=\t, domain=0:360]
  ({(1.08 + 0.03*cos(\t) + 0.02*sin(2*\t))*cos(\t)},
   {(0.80 + 0.02*sin(\t) - 0.03*cos(2*\t))*sin(\t)},
   {0.10*sin(4*\t) + 0.04*sin(\t)});

  \draw[thick,cyan]
    plot[smooth cycle, samples=180, variable=\t, domain=0:360]
    ({(1.08 + 0.03*cos(\t) + 0.02*sin(2*\t))*cos(\t)},
     {(0.80 + 0.02*sin(\t) - 0.03*cos(2*\t))*sin(\t)},
     {0.10*sin(4*\t) + 0.04*sin(\t)});

  \foreach \t in {-80,-60,...,120} {
    \pgfmathsetmacro{\xb}{(1.08 + 0.03*cos(\t) + 0.02*sin(2*\t))*cos(\t)}
    \pgfmathsetmacro{\yb}{(0.80 + 0.02*sin(\t) - 0.03*cos(2*\t))*sin(\t)}
    \pgfmathsetmacro{\zb}{0.10*sin(4*\t) + 0.04*sin(\t)}
    \draw[red] (Apex) -- (\xb,\yb,\zb);
  }

  \foreach \t in {120,140,...,260} {
    \pgfmathsetmacro{\xb}{(1.08 + 0.03*cos(\t) + 0.02*sin(2*\t))*cos(\t)}
    \pgfmathsetmacro{\yb}{(0.80 + 0.02*sin(\t) - 0.03*cos(2*\t))*sin(\t)}
    \pgfmathsetmacro{\zb}{0.10*sin(4*\t) + 0.04*sin(\t)}
    \draw[dashed, red] (Apex) -- (\xb,\yb,\zb);
  }

  \draw[thick,->] (0,0,0.0) -- (0,0,1.5);
  \draw[thick] (0,0,-0.27) -- (0,0,-0.5);

  \node at (0,0.8,1) {\textcolor{red}{$\CH^+: \quad R = 0$}};
  \node at (0.1,-1,-0.05) {\textcolor{cyan}{$\CB$}};
  \node at (0.1,-0.9,0.25) {\textcolor{blue}{$\tr = \tr_0$}};
  \node at (0,0.1,1.5) {{$t$}};
  
    \foreach \phi in {-60,-52,...,112} {
    \draw[blue,thick,opacity=0.28]
      plot[domain=1:3,samples=80,smooth,variable=\x]
      ({(3-\x)*cos(\phi)/3},{(3-\x)*sin(\phi)/3},{sqrt(\x*\x-1)/3});
  }
    \foreach \phi in {120,128,...,300} {
    \draw[blue,thick,opacity=0.28, dashed]
      plot[domain=1:3,samples=80,smooth,variable=\x]
      ({(3-\x)*cos(\phi)/3},{(3-\x)*sin(\phi)/3},{sqrt(\x*\x-1)/3});
  }

\end{tikzpicture}

\caption{We illustrate constant $\tr$ hypersurfaces within a causal diamond. In Gaussian null coordinates given in \eqref{eq:GN}, the future null horizon $\CH^+$ corresponds to $\tr=0$, whereas $\tr = \tr_0 > 0$ correspond to stretched horizons within the causal diamond. The past boundary of $\CH^+$ is given by the angle-dependent codimension-2 surface $\CB$. Note that as $\tr$ increases, the stretched horizon moves inwards into the causal diamond, indicating that $n = \dt\tr$ is inward-pointing. Each stretched horizon consists of worldlines of constant acceleration, and $\ell = \p_\tu$ is the clock along such trajectories.} \label{fig:GNC}

\end{figure}
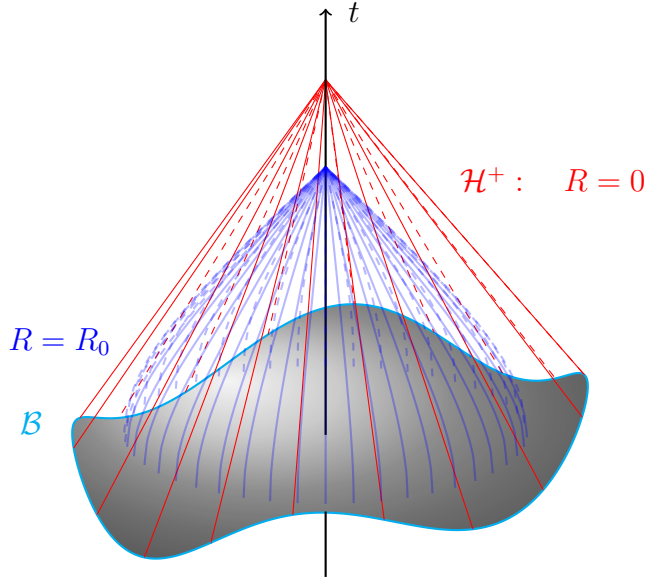

\subsection{Constructing the on-shell phase space}
\label{subsec:carroll}

In general, the gravitational phase space induced on a null hypersurface can be characterized in terms of a clean separation into spin-$0$, spin-$1$, and spin-$2$ degrees of freedom. The spin-$2$ sector describes the dynamical modes typically associated with gravitational radiation, such as gravitons. On the other hand, the spin-$0$ and spin-$1$ sectors of the theory are much less studied, since they play an important role only in the presence of boundaries. As we are focused on the boundary degrees of freedom, our primary interest will actually be precisely the spin-$0$ and spin-$1$  modes. Indeed, one of our main objectives is to construct a phase space which characterizes geometric fluctuations of a cut on a finite null hypersurface (see left diagram in Figure~\ref{fig:two-setups}). Therefore, it suffices for us to restrict our attention to metrics that can be written using Gaussian null coordinates and have no spin-$2$ modes. 

In $(d+2)$-dimensional spacetime, the line element can be written in Gaussian null coordinates as\footnote{We introduce here coordinates $x^\mu=(\tu,\tr,\vec\tx)$, where $\vec\tx \equiv (\tx^1,\ldots,\tx^d)$ label the transverse metric coordinates. Further, we will use lowercase Latin letters $i,j,\ldots$ to denote transverse components of the Carrollian line element, and reserve uppercase Latin letters $A,B,\ldots$ to denote transverse components of the (asymptotically) flat line element, which will appear in Section~\ref{sec:connection}. } 
\begin{align}\label{eq:GN}
\begin{split}
	\dt s^2 &= -2\ka(\tu,\vec\tx) \tr\,\dt \tu^2 + 2 \,\dt \tu\,\dt \tr - 4\tr \pi_i(\tu,\vec\tx) \,\dt \tu\,\dt \tx^i \\
	&\qquad + q_{ij}(\tu,\tr,\vec\tx) \dt \tx^i\,\dt \tx^j + \CO(\tr^2)  \\
	&= -2\ka(\tu,\vec\tx) \tr\,\dt \tu^2 + 2 \, \dt \tu\,\dt \tr - 4\tr \pi_i(\tu,\vec\tx) \,\dt \tu\,\dt \tx^i \\
	&\qquad + \varphi(\tu,\vec\tx)^{\frac{2}{d}} \Big( \g_{ij}(\vec\tx) + \tr \g_{ij}^\1(\tu,\vec\tx) \Big) \dt \tx^i\,\dt \tx^j + \CO(\tr^2) ,
\end{split}
\end{align}
where $\g_{ij}$ is the non-dynamical round metric on the transverse sphere, and in the second equality we have expanded the transverse metric $q_{ij}$ about $\tr = 0$, assuming that to leading order the transverse metric only has conformal time dependency, i.e.,
\begin{align}\label{trans-met}
	q_{ij}(\tu,\tr,\vec\tx) &= \varphi(\tu,\vec\tx)^{\frac{2}{d}}\Big( \g_{ij}(\vec\tx) + \tr \g_{ij}^\1(\tu,\vec\tx) \Big) + \CO(\tr^2) .
\end{align}
This assumption sets the shear,\footnote{The shear here refers to the $\partial_U$-shear, and should not be confused with the Bondi shear $C_{AB}$ discussed in Section~\ref{sec:connection} below. The Carrollian shear is the tracefree part of the Lie derivative of the null horizon metric along the clock vector field. Conversely, the Bondi shear parameterizes the large-$r$ expansion of an asymptotically flat metric, e.g., see \cite{Ciambelli:2025mex}.} which encodes the aforementioned spin-$2$ degrees of freedom, of the leading order transverse metric to zero. Thus, \eqref{trans-met} restricts our analysis to the boundary spin-0 and spin-1 modes. As is explained in Appendix~\ref{app:carroll-review}, the metric \eqref{eq:GN} is a specific instance of a more general class of Carrollian metrics.

The data specifying \eqref{eq:GN} consists of the inaffinity $\ka$, which is defined via the equation
\begin{align}\label{eq:kappa}
	\ell^\mu \nabla_\mu \ell^\nu = \ka \ell^\nu,
\end{align}
the Hájiček connection $\pi_i$, and the conformal factor $\varphi$, which controls the size of the leading transverse metric and can therefore be viewed as a dilaton or breathing mode.\footnote{The particular case where $\varphi$ is independent of the transverse coordinates $\vec\tx$ was studied in \cite{Bub:2024nan, Ciambelli:2025fbo}.} The metric \eqref{eq:GN} parameterizes a foliation of hypersurfaces as level sets of the nonnegative coordinate $\tr$, with the ``clock'' along any constant $\tr$ hypersurface being governed by the vector field $\ell = \partial_U$.\footnote{The metric \eqref{eq:GN} is closely related to the non-relativistic fluid metric studied in \cite{Bak:2024kzk}, with $\pi_i$ playing the role of the fluid velocity $v_i$ and the breathing mode $\varphi$ the role of the fluid potential $\phi$. As in \cite{Bak:2024kzk}, we will be interested in the case where the Hájiček connection is fully determined by the breathing mode. It would be interesting to study this connection more precisely in future work. \label{fn:fluid}} For later convenience, we also define the expansion scalar 
\begin{align}\label{eq:theta}
\begin{split}
	\th &\equiv  \p_U \log\varphi  .
\end{split}
\end{align}

Consider the hypersurface given by $\tr = \tr_0$. The normal one-form is given by $n \equiv \dt \tr = (0,1,\vec 0)$. Note that $n^2 = 2\ka \tr_0$, implying that the hypersurface is timelike when $\tr_0 > 0$ and becomes null when $\tr_0 = 0$. Now, given such a hypersurface, the line element \eqref{eq:GN} pulls back to be the induced line element
\begin{align}
\begin{split}
	\dt s^2\big|_{\tr = \tr_0} &= -2\ka(\tu,\vec\tx) \tr_0\,\dt \tu^2 - 4\tr_0 \pi_i(\tu,\vec\tx) \,\dt \tu\,\dt \tx^i +\\
	&\qquad \varphi(\tu,\vec\tx)^{\frac{2}{d}} \Big( \g_{ij}(\vec\tx) + \tr_0 \g_{ij}^\1(\tu,\vec\tx) \Big) \dt \tx^i\,\dt \tx^j + \CO(\tr_0^2) .
\end{split}
\end{align}
As we mentioned above, for $\tr_0 > 0$, this line element describes a timelike hypersurface, with $\sqrt{2\ka\tr_0}$ effectively playing the role of the speed of light. Thus, the limit $\tr_0\to 0$ is a Carrollian limit \cite{Donnay:2019jiz}, in which case the induced line element becomes null:
\begin{align}\label{eq:ds20}
	\dt s^2\big|_{\tr = 0} &= \varphi(\tu,\vec\tx)^{\frac{2}{d}} \g_{ij}(\vec\tx)  \dt \tx^i\,\dt \tx^j.
\end{align}
It is clear that the metric degenerates in this limit as only the transverse metric components remain nonzero. As this is the regime we are interested in, we are justified in neglecting all the higher order $\tr$ terms in \eqref{eq:GN}.

Recall that the pre-symplectic potential for Einstein--Hilbert gravity is given by (e.g., see \cite{Carroll:2004st})
\begin{align}\label{eq:symp-pot}
	\Th &=  \frac{1}{16\pi G} \int_\CH \dt \Sigma_\sigma \big( g^{\sigma\mu} \nabla^\nu \d g_{\mu\nu} - g^{\mu\nu} \nabla^\sigma \d g_{\mu\nu} \big) ,
\end{align}
where $\dt\Sigma_\s$ is the induced area element on $\CH^{+}$ given by\footnote{The negative sign in \eqref{eq:induced-area} follows from the induced area element being defined to be outward-pointing, while the normal one-form $\dt R$ is inward-pointing. This is because increasing $\tr$ corresponds to moving deeper into the interior of the causal diamond, as shown in Figure~\ref{fig:GNC}.}
\begin{align}\label{eq:induced-area}
	\dt\Sigma_\s = - \d^\tr_\s \varphi\sqrt{\g}\,\dt \tu\,\dt^d\tx, \qquad \sqrt \g \equiv \sqrt{\det\g_{ij}} .
\end{align}
Our phase space is described by the line element \eqref{eq:GN}, and the dynamical fields consist of $\varphi,\ka,\pi_i$, and $\gamma^{(1)}_{ij}$. As we mentioned earlier, $\g_{ij}$ is non-dynamical, i.e., $\delta\gamma_{ij}=0$, so that only the conformal factor in the degenerate metric \eqref{eq:ds20} on $\CH^{+}$ is allowed to vary.  

We are interested in evaluating the pre-symplectic potential on the null hypersurface $\tr = 0$. Substituting our metric \eqref{eq:GN} into \eqref{eq:symp-pot} with $\tr = 0$, so that $\CH = \CH^+$, we get the pre-symplectic potential in Gaussian null coordinates to be
\begin{align}
\begin{split}
	\Th &= \frac{1}{8\pi G} \int_{\CH^+} \dt \tu\,\dt^d \tx\,\sqrt \g \, \varphi \bigg( \d \ka + \frac{1}{\varphi} \p_\tu \d\varphi - \frac{d-1}{d} \frac{\p_\tu\varphi}{\varphi^2}\d\varphi \bigg)  \\
	&=  - \frac{1}{8\pi G} \int_{\CH^+} \dt U\,\dt^d \tx\,\sqrt{\g}  \bigg( \ka \d \varphi + \frac{d-1}{d} \th\d\varphi - \d( \varphi\ka + \p_U \varphi ) \bigg) \\
	&=  - \frac{1}{8\pi G} \int_{\CH^+} \dt U\,\dt^d \tx\,\sqrt{\g}  \bigg( \mu \d \varphi  - \d( \varphi\ka + \p_U \varphi ) \bigg)  , 
\end{split}
\end{align}
where in the second line we used \eqref{eq:theta}, and in the last line we defined
\begin{equation}
\mu \equiv \ka + \frac{d-1}{d} \th.
\label{eq:mu}
\end{equation} 
This combination, introduced in \cite{Hopfmuller:2018fni}, plays an important role in the null quantization \cite{Ciambelli:2024swv}.

The pre-symplectic form is obtained by taking another exterior derivative in phase space, so that
\begin{align}\label{eq:symp-form}
\begin{split}
	\O \equiv \d\Th = -\frac{1}{8\pi G} \int_{\CH^+} \dt \tu\,\dt^d \tx\,\sqrt \g \, \d \mu \wedge \d\varphi ,
\end{split}
\end{align}
where we used the identity $\d^2 = 0$. It is worth noting that the Hájiček connection $\pi_i$, which would have been the independent spin-$1$ degree of freedom, does not appear in \eqref{eq:symp-form}. This is because our choice of metric has fixed the clock vector field $\ell = \pa_U$ to be non-dynamical, with $\ell^i=0$. As the transverse part of the clock vector field is canonically conjugate to $\pi_i$ \cite{Chandrasekaran:2018aop,Chandrasekaran:2023vzb,Odak:2023pga,Ciambelli:2023mir,Klinger:2025tvg}, taking $\delta \ell^i = 0$ forces $\delta \ell^i\wedge \delta \pi_i = 0$. However, as we shall see below, once we impose the equations of motion, the Hájiček connection $\pi_i$ itself is \emph{not} equal to zero; rather, it is simply determined entirely by the dynamical spin-$0$ data. This is consistent with the analysis for a non-relativistic fluid performed in \cite{Bak:2024kzk}, where the fluid velocity is determined by the transverse derivatives of a scalar potential (see Footnote~\ref{fn:fluid}).

Let us now turn to imposing the Einstein equations on $\CH^{+}$. Famously, the intrinsic piece of the Einstein equations projected on a timelike or null hypersurface gives rise to the Raychaudhuri and the momentum constraint equations. For the case where the hypersurface is null, which is the case of interest for us, the momentum constraint becomes the Damour equation \cite{Damour:1985cm}. Because we have not imposed such equations of motion yet, our pre-symplectic form \eqref{eq:symp-form} is still \emph{kinematical}. We would now like to impose the constraint equations and derive the \emph{physical} pre-symplectic form. In particular, for a metric with no shear, which is the case for our line element \eqref{eq:GN}, the Raychaudhuri and Damour equations are respectively\footnote{The Damour equation explicitly depends on the choice of horizontal sub-bundle, and thus on the Ehresmann connection $k$ dual to $\ell$. Here, we chose $k =\rd U$ so that $\rd k=0$.} 
\begin{align}
	(\p_\tu +\th)\th = \mu\th \label{eq:ray} \\
	(\p_\tu + \th) \pi_i = \p_i\mu \label{eq:dam} ,
\end{align}
where both equations are implicitly defined on $\CH^+$. These are precisely the $UU$ and $Ui$ components of the Einstein equations associated to the metric \eqref{eq:GN} for $R=0$.

It is possible to solve both \eqref{eq:ray} and \eqref{eq:dam} exactly, and the details are found in Appendix~\ref{app:RayDam}. For the purposes of the present analysis, however, it is more direct to enforce the equations \eqref{eq:ray} and \eqref{eq:dam} as relations between phase space variables and simplify the resulting symplectic form.\footnote{Since $\pi_i$ does not appear in \eqref{eq:symp-form}, we do not need to use the Damour equation \eqref{eq:dam}. However, there is nontrivial spin-1 data as the line element \eqref{eq:GN} explicitly contains $\pi_i$; it just happens to be dynamically determined via the spin-0 degrees of freedom (see \eqref{eq:data} below). This differs from the previous work \cite{Bub:2024nan, Ciambelli:2025fbo}, where the spin-0 sector was assumed to be spherically symmetric, and so the spin-1 sector vanishes completely.} Assuming that $\th$ is nowhere vanishing, we can rewrite \eqref{eq:ray} as
\begin{align}\label{mu-onshell}
	\mu = \th + \p_\tu\log|\th|.
\end{align}
Substituting this back into \eqref{eq:symp-form}, we get
\begin{align}\label{eq:symp-form2}
\begin{split}
	\O &= -\frac{1}{8\pi G} \int_{\CH^+} \dt \tu\,\dt^d \tx\,\sqrt \g \, \big( \d \th + \p_U \d\log|\th|  \big) \wedge \d\varphi \\
	&= -\frac{1}{8\pi G} \int_{\p\CH^+}\dt^d \tx\,\sqrt \g \, \d\log|\th| \wedge \d\varphi -\frac{1}{8\pi G}  \int_{\CH^+} \dt \tu\,\dt^d \tx\,\sqrt \g \, \bigg( \d \th \wedge \d\varphi - \frac{\d\th}{\th} \wedge \d(\th\varphi) \bigg) \\
	&= \frac{1}{8\pi G} \int_{\CB} \dt^d \tx\,\sqrt \g \, \d\log |\th| \wedge \d\varphi ,
\end{split}
\end{align}
where in the second equality we integrated by parts and then used \eqref{eq:theta}; and in the last equality we used the antisymmetry of the wedge product and also noted the future boundary of $\CH^{+}$ is a caustic where $\d\varphi$ vanishes, and the past boundary is the codimension-2 surface $\CB$ (see Figure~\ref{fig:GNC}). Because $\O$ in \eqref{eq:symp-form2} is already in Darboux form, it is manifestly invertible and therefore is not just the pre-symplectic form, but the \emph{symplectic} form. For the specific case of four spacetime dimensions, we have $d=2$. In this case, we can choose $\tx^i = (\tz,\tbz)$ to be the complex coordinates on the sphere. The round metric $\g_{ij}$ is then given by $\g_{\tz\tbz} = \frac{2}{(1+\tz\tbz)^2}$ and vanishes otherwise, and the symplectic form \eqref{eq:symp-form2} becomes
\begin{align}\label{4d-symp-form}
\begin{split}
	\O &= \frac{1}{8\pi G} \int_{\CB} \dt^2\tz \,\g_{\tz\tbz} \, \d\log|\th| \wedge \d\varphi ,  \qquad \dt^2\tz \equiv -i \,\dt\tz\,\dt\tbz .
\end{split}
\end{align}

\subsection{A finite Minkowski subregion}\label{subsec:mink}

In the previous subsection, we constructed the constrained symplectic form associated to shape deformations of a null hypersurface in an arbitrary background. We would now like to specialize to the particular case where our null hypersurface lives in four-dimensional Minkowski spacetime. Even though this is a special case of our more general analysis, it is instructive to understand exactly what the expansion scalar $\th$ and dilaton $\varphi$ correspond to in a finite subregion in Minkowski spacetime. These observations will prove crucial in constructing the isomorphism between the phase space of cuts on a finite null hypersurface and that of asymptotically flat spacetimes. 

The line element in four-dimensional Minkowski spacetime is given by
\begin{align}\label{eq:mink}
	\dt s^2 = -\dt u^2 - 2 \,\dt u\,\dt r + r^2\g_{z\bz} \, \dt z\,\dt\bz,
\end{align}
where $u$ is the retarded null time, $r$ the usual radial coordinate, and $(z,\bz)$ complex stereographic coordinates on the transverse sphere. Note that future null infinity $\SI^+$ is reached by taking $r\to\infty$ while keeping $u$ fixed. We determine in Appendix~\ref{app:diffeo} that there exists a coordinate map that transforms the Minkowski metric \eqref{eq:mink} involving coordinates $(u,r,z,\bz)$ to the neighborhood of the $\tr=0$ null hypersurface in Gaussian null coordinates, which is given by \eqref{eq:GN} with coordinates $(\tu,\tr,\tz,\tbz)$. This coordinate map is given by \eqref{diffeo2} to be
\begin{align}\label{diffeo-map}
\begin{split}
	u(\tu,\tr,\tz,\tbz) &= L_0 - 2\Phi + \frac{2\g^{\tz\tbz} \p_\tz\Phi\p_\tbz\Phi}{\Phi^2\p_\tu\Phi}\tr - \frac{2 \big(\g^{\tz\tbz}\p_\tz\Phi \p_\tbz\Phi\big)^2}{\Phi^5(\p_\tu\Phi)^2 }\tr^2 + \cdots \\
	r(\tu,\tr,\tz,\tbz) &=  \Phi + \frac{\Phi^2 - 2\g^{\tz\tbz}\p_\tz\Phi\p_\tbz\Phi}{2\Phi^2\p_\tu\Phi} \tr + \frac{\g^{\tz\tbz} \p_\tz\Phi\p_\tbz \Phi \big( \Phi^2 + 2\g^{\tz\tbz}\p_\tz\Phi\p_\tbz\Phi \big) }{2\Phi^5 (\p_\tu\Phi)^2} \tr^2 + \cdots   \\
	z(\tu,\tr,\tz,\tbz) &= \tz - \frac{\g^{\tz\tbz} \p_\tbz\Phi}{\Phi^2\p_\tu\Phi} \tr - \frac{\G^{\tz}_{\tz\tz}\big(\g^{\tz\tbz}\p_\tbz \Phi\big)^2}{2\Phi^4(\p_\tu\Phi)^2} \tr^2 + \cdots ,
\end{split}
\end{align}
where $\cdots$ indicate further subleading $\CO(\tr^3)$ terms, and $L_0$ is an arbitrary spacetime constant determined by boundary conditions (see \eqref{mink-null} below). Substituting this into the Minkowski line element \eqref{eq:mink}, we get precisely the Gaussian null line element \eqref{eq:GN}, with the data given in \eqref{asy-data} to be
\begin{align}\label{eq:data}
\begin{split}
	\varphi &= \Phi^2 \\
	\ka &= \frac{\p_\tu^2 \Phi}{\p_\tu \Phi} = \p_\tu\log |\p_\tu \Phi |  \\
	\pi_\tz &= \frac{\p_\tu\p_\tz \Phi}{\p_\tu \Phi} - \frac{\p_\tz\Phi}{\Phi} = \p_\tz  \log \big|\p_\tu \log \Phi \big|  \\
	q_{\tz\tz} &=  \frac{2\tr}{\Phi\p_\tu\Phi} \big[ 2 (\p_\tz \Phi)^2 - \Phi  D_\tz^2 \Phi  \big] \\
	q_{\tz\tbz} &=  \g_{\tz\tbz}\Phi^2 + \frac{2\tr}{\p_\tu\Phi} \bigg( \frac{2\p_\tz\Phi\p_\tbz\Phi}{\Phi} - \p_\tz\p_\tbz\Phi  \bigg) ,
\end{split}
\end{align}
Importantly, notice that all the data in \eqref{eq:data} are determined by a single spin-0 scalar mode $\Phi$. Notice though that because $\th$, defined in \eqref{eq:theta}, can be written in terms of $\Phi$ as
\begin{align}\label{th-Phi}
	\th = 2\p_\tu \log\Phi,
\end{align}
we can also more compactly rewrite $\pi_\tz$ as
\begin{align}
	\pi_\tz = \p_\tz \log|\th|.
\end{align}

As we are studying the neighborhood of the $\tr=0$ null hypersurface in Gaussian null coordinates, we would like to determine how to parametrize this hypersurface in Minkowski coordinates. For $\tr=0$, the coordinate map \eqref{diffeo-map} becomes
\begin{align} \label{R0-map}
\begin{split}
	u = L_0 - 2\Phi(\tu,\tz,\tbz) , \qquad r = \Phi(\tu,\tz,\tbz) , \qquad z = \tz.
\end{split}
\end{align}
This implies that we have the equality
\begin{align}\label{mink-null}
	u + 2r = L_0 ,
\end{align}
so this is describing, for \emph{any} choice of $\Phi$, the null hypersurface $\CH^+$ intersecting the $t=0$ hyperplane at radius $r=L_0$. However, in our analysis we are only interested in a subregion of $\CH^+$, namely the region consisting of ingoing null rays from a cut $\CB$ on the hypersurface, as depicted by red lines in Figure~\ref{fig:GNC}. Thus, we demand that the past boundary $\CB$ is located at $\tu = \tu_\CB \to -\infty$ and $\tr=0$. Letting
\begin{align}\label{bdy-cond}
	\lim_{\tu \to -\infty} \Phi(\tu, \tz,\tbz) = L(\tz,\tbz),
\end{align}
we will now use this boundary condition to solve for $\Phi$ as a function of $\ka$. 

Because our symplectic form only depends on the corner, we can perform a small gauge-fixing so that the inaffinity is time-independent, i.e., $\ka(\tu,\tz,\tbz) \equiv \ka(\tz,\tbz)$.\footnote{Note that for more general spacetimes, where the symplectic form lives on the entire horizon $\CH^+$, assuming the inaffinity $\ka$ is time-independent is indeed a restriction. For the reduced phase space given by the symplectic form \eqref{4d-symp-form} however, we can fix $\ka$ to be time-independent using a field redefinition that preserves the symplectic structure. \label{fn:small-gauge} } In this case, we can integrate the equation for $\ka$ given in \eqref{eq:data} to obtain
\begin{align}\label{k-constant}
	\log|\p_\tu\Phi(\tu,\tz,\tbz)| = \ka(\tz,\tbz) \tu + \a(\tz,\tbz),
\end{align}
where $\a(\tz,\tbz)$ is an integration constant. Exponentiating both sides of \eqref{k-constant}, we get
\begin{align}\label{D-Phi1}
\begin{split}
	| \p_\tu \Phi(\tu,\tz,\tbz) | = e^{\ka(\tz,\tbz)\tu + \a(\tz,\tbz)} .
\end{split}
\end{align}
Because the right-hand side is positive for any finite $\tu$, this means that, assuming $\Phi$ is smooth, $\p_\tu\Phi$ is either always positive or always negative. If $\p_\tu\Phi > 0$, then it is obvious $\Phi$ will diverge. Hence, we conclude that we need $\p_\tu\Phi$ to always be negative in \eqref{D-Phi1}, allowing us to rewrite it as
\begin{align}\label{phi-soln}
\begin{split}
	&\p_\tu\Phi(\tu,\tz,\tbz) = - e^{\ka(\tz,\tbz)\tu + \a(\tz,\tbz)} \\
	\implies\quad  & \Phi(\tu,\tz,\tbz) = L(\tz,\tbz) - \frac{1}{\ka(\tz,\tbz)} e^{\ka(\tz,\tbz) \tu + \a(\tz,\tbz)} ,
\end{split}
\end{align}
where the integration constant $L(\tz,\tbz)$ is fixed using the boundary condition \eqref{bdy-cond}. 

We can now use \eqref{phi-soln} to determine the $\tu$ coordinate for the top tip of the diamond as a function of the angles, which we denote as $\tu_+$. This occurs when $r=0$ on $\CH^+$, which implies that
\begin{align}
\begin{split}
	& \Phi(\tu_+ ,\tz,\tbz) = L(\tz,\tbz) - \frac{1}{\ka(\tz,\tbz)} e^{\ka(\tz,\tbz) \tu_+ + \a(\tz,\tbz)} = 0 \\
	\implies\quad & \tu_+ = \frac{1}{\ka(\tz,\tbz)} \Big[ \log\big( \ka(\tz,\tbz) L(\tz,\tbz) \big) - \a(\tz,\tbz) \Big] .
\end{split}
\end{align}
Note that because all the points on $\CB$ correspond to $\tu_\CB \to -\infty$, even though they are at different $t$ values in Minkowski coordinates, the $\tu_+(\tz,\tbz)$ values for each angle is different despite the fact all the null rays meet at the same point. Finally, we may assume without loss of generality that the cut $\CB$ lives in the upper half space and demand $t|_\CB \geq 0$. Noting that
\begin{align}
	t = u + r = L_0 - \Phi(\tu,\tz,\tbz),
\end{align}
it follows that requiring $t \geq 0$ on $\CB$ implies
\begin{align}
\begin{split}
	0 \leq \lim_{\tu \to -\infty} t(\tu, R=0,\tz,\tbz) = L_0 - L(\tz,\tbz) \quad\implies\quad L_0 \geq L(\tz,\tbz).
\end{split}
\end{align}

In the last part of this subsection, we will use \eqref{phi-soln} to rewrite the constrained symplectic form \eqref{4d-symp-form}. Since Minkowski spacetime is a solution to the Einstein equations, the Gaussian null metric with data \eqref{eq:data} automatically satisfies both the Raychaudhuri and Damour equations. Substituting \eqref{th-Phi} into \eqref{4d-symp-form}, we get
\begin{align}\label{4d-symp2}
\begin{split}
	\O &= \frac{1}{8\pi G} \int_{\CB} \dt^2\tz \,\g_{\tz\tbz} \, \d \big( \log|2\p_\tu\log\Phi| \big) \wedge \d\varphi \\
	&=  \frac{1}{8\pi G} \int_{\CB} \dt^2\tz \,\g_{\tz\tbz} \, \frac{\Phi}{\p_\tu\Phi} \d \bigg( \frac{\p_\tu\Phi}{\Phi} \bigg) \wedge \d\varphi \\
	&=  \frac{1}{8\pi G} \int_{\CB} \dt^2\tz \,\g_{\tz\tbz} \, \frac{\d ( \p_\tu\Phi) }{\p_\tu\Phi} \wedge \d\varphi \\
	&=  \frac{1}{8\pi G} \int_{\CB} \dt^2\tz \,\g_{\tz\tbz} \d \log|\p_\tu\Phi|  \wedge \d\varphi ,
\end{split}
\end{align}
where in the second equality we used the chain rule for $\d$, and in the third equality we used the fact the wedge product $\wedge$ is antisymmetric and $\d\varphi \sim \d\Phi$. Furthermore, as was demanded above \eqref{bdy-cond}, the past boundary $\CB$ corresponds to $\tu = \tu_\CB \to -\infty$ and $\tr=0$. It follows from \eqref{phi-soln} that
\begin{align}\label{Phi-B1}
	\varphi\big|_\CB = \Phi^2\big|_{\CB} = L(\tz,\tbz)^2.
\end{align}
Substituting \eqref{k-constant} and \eqref{Phi-B1} into \eqref{4d-symp2}, we get
\begin{align}
	\O = \frac{1}{4\pi G} \int_\CB \dt^2\tz\,\g_{\tz\tbz}  \d \big( \ka \tu_\CB + \a \big)  \wedge (L \d L) .
\end{align}
Notice that $\alpha$ is the finite part of $\log \theta$ on the corner, since $\tu_\CB \to -\infty$ diverges there. Indeed, this divergence may appear to cause the symplectic form to be divergent as well. However, if we assume that $\ka(\tz,\tbz)$ only depends on $L(\tz,\tbz)$, which is simply parametrizing the size of $\CB$ in the $(\tz,\tbz)$ direction, the antisymmetry of the wedge product $\wedge$ eliminates the divergence coming from $\tu_\CB$, resulting in the final  symplectic form\footnote{An alternative approach would be to regulate the divergent $U_\CB$-limit, and was explored in \cite{Fransen:2025npa}.} 
\begin{align}\label{4d-symp3}
\begin{split}
	\O &=  \frac{1}{4\pi G} \int_{\CB} \dt^2\tz \,\g_{\tz\tbz} \, \d (\a L )  \wedge \d L .
\end{split}
\end{align}
Inverting the symplectic form \eqref{4d-symp3}, we get the bracket
\begin{align}\label{bracket1}
\begin{split}
	L(z,\bz) \{ \a(z,\bz) , L(w,\bw) \} = - 4\pi G \g^{z\bz} \d^2(z-w) ,
\end{split}
\end{align}
where we have used the fact that $z=\tz$ on the horizon $\CH^+$ (and hence also on $\CB$) according to \eqref{R0-map}.

For later convenience, it is useful to recast the bracket \eqref{bracket1} in a slightly different form. First, we want to isolate the dynamical part of the field $L(z,\bz)$. This can be accomplished by choosing fixed background values of the phase space variables and considering fluctuations. One way to do this is to choose a probability distribution $\rho[L,\alpha]$ on the phase space and expand general phase space points around the mean of this distribution, which is given by
\begin{align}\label{L-mean}
\begin{split}
	\bar L(z,\bz) \equiv \int \dt L \,\dt \a \, \rho[L,\alpha] L(z,\bz) .
\end{split}
\end{align}
We note that $\bar{L}(z,\bz)$ is a spacetime function but a phase space constant since it is obtained by integrating over phase space. We view $\bar L(z,\bz)$ as the average size of the cut $\CB$ in the direction $(z,\bz)$, and interpret it as encoding the non-fluctuating part of the mode $L(z,\bz)$. Then, we can isolate the dynamical part of $L(z,\bz)$ by subtracting off $\bar L$ and define
\begin{align}\label{eps-def}
\begin{split}
	\e(z,\bz) \equiv L(z,\bz) - \bar L(z,\bz).
\end{split}
\end{align}
Substituting this into the bracket \eqref{bracket1} and noting that $\{\bar L(z,\bz) , \cdot \} = 0$ since $\bar L(z,\bz)$ is a phase space constant, we find
\begin{align}\label{final-bracket}
\begin{split}
	\big\{ \a(z,\bz) ,  \e(w,\bw)  \big\} = - \frac{4\pi G}{L(z,\bz)} \g^{z\bz} \d^2(z-w) , 
\end{split}
\end{align}
where we have assumed that $L$ is nowhere vanishing when dividing it. This is precisely the angle-dependent generalization of the analysis in \cite{Ciambelli:2025fbo}. 
 
It may be surprising that we obtain a non-vanishing symplectic form even though we are studying a subregion of \emph{pure} Minkowski spacetime, where there are no fluctuating degrees of freedom in the metric. Nevertheless, a useful point of view advanced in \cite{Ciambelli:2024vhy} is that the dynamical modes are precisely the \emph{large} diffeomorphism modes acting on $\CH^+$ in the map \eqref{diffeo-map}, as changing such modes changes the boundary data \eqref{eq:data}. This perspective becomes clear when we solved for $\Phi$ in \eqref{phi-soln}, revealing that the integration constants $L(\tz,\tbz)$ and $\a(\tz,\tbz)$ are determined by the boundary conditions that we impose in the original Minkowski spacetime, and these are the quantities appearing in the symplectic form \eqref{4d-symp3}.

We conclude this subsection by noting that the symplectic form completely localizes to the codimension-2 corner. This is exactly what happens in the infrared phase space of asymptotically flat spacetimes, which we shall now review in the next section. We will then construct an isomorphism between the phase spaces in these two different physical setups. This generalizes the results of \cite{Ciambelli:2025fbo}, which constructed the isomorphism for the special case where spherical symmetry is imposed.

\section{Connecting to asymptotically flat spacetimes}\label{sec:connection}

In the previous section, we constructed the phase space for the future horizon of a causal diamond embedded within Minkowski spacetime. This phase space has local coordinates $(L,\alpha)$ associated with the length and the initial time offset, respectively. They are related to the remaining gravitational degrees of freedom through \eqref{eq:data}, which encode the details of this particular solution to the Einstein equations. 

We would now like to relate this phase space to that of asymptotically flat spacetimes. We will in Section~\ref{subsec:afs} give a quick review of such spacetimes and their associated phase space, which describes the dynamical modes of asymptotically flat gravity. In particular, we are interested in specializing to the soft sector of this phase space, which consists of the leading soft graviton and supertranslation Goldstone modes $N,C$, respectively. We will then in Section~\ref{subsec:iso} construct the explicit isomorphism preserving the symplectic bracket between this phase space and that of the causal diamond. As we shall see, the area fluctuations for the asymptotically flat spacetimes are controlled by the soft graviton $N(z,\bz)$, which is identified with the radial fluctuation mode $\e(z,\bz)$ controlling the size of the transverse cut for a finite causal diamond. We can then use the symplectic structure to identify the Goldstone $C(z,\bz)$ with the null time offset mode $\a(z,\bz)$ labeling the cut $\CB$.

\subsection{Brief review of asymptotically flat spacetimes}\label{subsec:afs}

To construct the phase space for an asymptotically flat spacetime, it is convenient to work in Bondi gauge, where the line element near $\SI^+$ is given by \cite{Bondi:1962px, Sachs:1962zza}
\begin{align}\label{eq:bondi}
\begin{split}
	\dt s^2 &= - \dt u^2 - 2 \, \dt u\,\dt r + 2 r^2\g_{z\bz}\,\dt z\,\dt \bz + \frac{2m_B}{r}\,\dt u^2 + r C_{zz} \, \dt z^2 + r C_{\bz\bz} \, \dt \bz^2 + \cdots,
\end{split}
\end{align}
where $u = t-r$ is the retarded null time, $m_B$ the Bondi mass aspect, $C_{zz}$ the shear, $\g_{z\bz} = \frac{2}{(1+z\bz)^2}$ is again the round metric, and $\cdots$ indicates further terms subleading in $r$ that are determined by the Einstein equations. The $r \to \infty$ limit corresponds to approaching $\SI^+$, while subsequently taking $u \to \pm \infty$ corresponds to the future and past boundaries $\SI^+_\pm$, respectively. The asymptotic data specifying the spacetime consists of $m_B$ and $C_{zz}$. Moreover, the vacuum Einstein equations impose a constraint on $m_B$, namely
\begin{align}\label{eq:mBconstraint}
	\p_u m_B = \frac{1}{4} ( D_z^2 N^{zz} + D_\bz^2 N^{\bz\bz} ) - \frac{1}{4} N_{zz} N^{zz} ,
\end{align}
where $N_{zz} \equiv \p_u C_{zz}$ is the news tensor measuring the gravitational radiation passing through $\SI^+$, and $D_z$ is the $\g$-covariant derivative. 

The symplectic form associated to the metric \eqref{eq:bondi} can be determined using the standard covariant phase space method, where we first obtain the symplectic form via \eqref{eq:symp-pot} and then take another exterior derivative in phase space. This results in the symplectic form \cite{Ashtekar:1981bq}
\begin{align}\label{afg-symp-form}
	\O_{\AFG} = - \frac{1}{16\pi G} \int_{\SI^+} \dt u\,\dt^2z \g^{z\bz} \d C_{zz} \wedge \d N_{\bz\bz} .
\end{align}
As we are only interested in the leading soft behavior, we restrict ourselves to the phase space associated to the leading soft graviton and supertranslation Goldstone modes, which are defined respectively as
\begin{align} \label{CN-def}
	D_z^2 N(z,\bz) \equiv \int_{-\infty}^\infty \dt u\, N_{zz}(u,z,\bz), \qquad -2 D_z^2 C(z,\bz) \equiv C_{zz}(u,z,\bz) \big|_{\SI^+_-} .
\end{align} 
Notice that both $N$ (the leading soft graviton mode) and $C$ (the supertranslation Goldstone mode) are scalar modes. Equivalently, the soft graviton and Goldstone modes in \eqref{CN-def} are captured by the memory shear profile \cite{Strominger:2014pwa, He:2024vlp}
\begin{align}\label{soft-shear}
\begin{split}
	C_{zz}(u,z,\bz) &= D_z^2 N(z,\bz) \th(u-u_s) - 2 D_z^2 C(z,\bz),
\end{split}
\end{align}
where $\th(u-u_s)$ is the Heaviside step function, and $u_s$ is the location of the shockwave implementing the gravitational memory effect by shifting the shear $C_{zz}$. The symplectic form involving these soft degrees of freedom is then given by \cite{He:2014laa, He:2024vlp, Ciambelli:2025fbo}
\begin{align} \label{eq:soft-symp}
\begin{split}
	\O_\AFG^\soft = \frac{1}{8\pi G} \int_{S^2} \dt^2z\, \g^{z\bz} D_z^2 \d C \wedge D_\bz^2 \d N,
\end{split}
\end{align}
where $S^2$ is the celestial sphere located at $\SI^+_-$. As already remarked, both this symplectic form and the one given in \eqref{4d-symp-form} are defined on a codimension-2 corner. Inverting the symplectic form \eqref{eq:soft-symp}, we obtain the bracket between the soft graviton and Goldstone mode, which is given by
\begin{align} \label{eq:softBracket}
\begin{split}
	&\big\{ D_z^2 C(z,\bz) , D_\bw^2 N(w,\bw) \big\} = -8\pi G \g_{z\bz} \d^2(z-w) \\
	\implies\quad & \big\{ C(z,\bz) , N(w,\bw) \big\} = -8\pi G S \log|z-w|^2 , \qquad S = \frac{(z-w)(\bz-\bw)}{(1+z\bz)(1+w\bw)},
\end{split}
\end{align}
where the implication follows from integrating over the transverse directions. This in turn implies
\begin{align}\label{soft-comm0}
	\big\{ C(z,\bz), D_w^2 D_\bw^2 N(w,\bw) \big\} = -8\pi G \g_{z\bz}\d^2(z-w).
\end{align}

It is useful for later purposes to rewrite \eqref{soft-comm0} using the transverse Laplacian. Noting that $D_z^2 D_\bz^2 N = D_\bz^2 D_z^2 N$, it follows
\begin{align}
	\Box N(z,\bz) \equiv (D^z D_z + D^\bz D_\bz ) N(z,\bz) = 2 D^z D_z N(z,\bz),
\end{align}
which then implies
\begin{align}\label{2-deriv}
\begin{split}
	(\g^{z\bz})^2 D_z^2 D_\bz^2 N(z,\bz) = \frac{1}{4} \Box ( \Box + 2) N(z,\bz).
\end{split}
\end{align}
Using \eqref{2-deriv}, we can then rewrite \eqref{soft-comm0} as
\begin{align}\label{soft-comm}
\begin{split}
	\big\{ C(z,\bz) , \Box(\Box+2) N(w,\bw) \big\} = -32\pi G \g^{z\bz} \d^2(z-w).
\end{split}
\end{align}

\subsection{The phase space isomorphism}\label{subsec:iso}

In this subsection, we will determine the relationship between the phase space of cuts on a finite null hypersurface in Minkowski spacetime described in Section~\ref{subsec:mink} and that of the soft sector of asymptotically flat spacetimes. In particular, we will construct an invertible map $\Psi$ between the two phase spaces that preserves the symplectic structure, i.e., a symplectomorphism. To arrive at this map, we first note that the degrees of freedom for the finite subregion are located on the null horizon $\CH^+$, which is located at $\tr=0$. Since $z=\tz$ according to \eqref{diffeo-map}, this means we can use the coordinate $(\tu,z,\bz)$ to describe all the fields on $\CH^+$ with the simple replacement $\tz \to z$, e.g.,
\begin{align}
	\Phi(\tu,\tz,\tbz) = \Phi(\tu,z,\bz).
\end{align} 
We will henceforth make this restriction.

Schematically, the construction of our phase space identification proceeds in two steps. First, we take advantage of the fact that these two phase spaces are fundamentally describing a common physical process, namely fluctuations of the area of a codimension-2 surface. This perspective allows us to construct a natural, geometric identification between the phase space variable $\e(z,\bz)$, which encodes area fluctuations of the corner of the causal diamond ,and $N(z,\bz)$, which sources area fluctuations of the celestial sphere in asymptotically flat spacetimes through \eqref{eq:mBconstraint}. After making this identification, we can then utilize the symplectic structure to guide our identification between the remaining conjugate variables. Given the map $\Psi$, we can translate between the physics of area fluctuations in the causal diamond and that of the asymptotically flat spacetime, providing a useful duality frame for doing computations in either regime. 

Following the plan described above, we first aim to relate the soft graviton mode $N$ to the radial fluctuation mode $\e$ of a finite null hypersurface. For any fixed $(z,\bz)$, the induced radial coordinate on a null hypersurface is given by \cite{Kapec:2016aqd, Ciambelli:2024swv}
\begin{align}\label{a-def}
	\p_u \big( r(u,z,\bz)^2 \big) = 2 m_B(u,z,\bz) - r(u,z,\bz).
\end{align}
For the case of the finite null horizon $\CH^+$ in Minkowski spacetime, the Bondi mass aspect vanishes. Therefore, along $\CH^+$, \eqref{a-def} reduces to
\begin{align}
\begin{split}
	\p_u \big( r_{\CH^+}(u,z,\bz)^2 \big) = - r_{\CH^+}(u,z,\bz) \implies\quad & r_{\CH^+}(u,z,\bz) = \frac{1}{2} (L_0 - u) ,
\end{split}
\end{align}
where we determined the integration constant in the implication by requiring $r_{\CH^+}(L_0) = 0$ at the top tip of the causal diamond, which is where the caustic is located. Denoting the local area element associated to any fixed $u$ as $a_{\CH^+}(u,z,\bz)$, we then have
\begin{align}
\begin{split}
	a_{\CH^+}(u,z,\bz) \equiv \g_{z\bz} r_{\CH^+}(u,z,\bz)^2 = \frac{1}{4}\g_{z\bz}(L_0 - u)^2.
\end{split}
\end{align}
It follows the rate of change of the local area element along $\CH^+$ is
\begin{align}
\begin{split}
	\p_u a_{\CH^+}(u,z,\bz) = - \frac{1}{2} \g_{z\bz}(L_0 - u).
\end{split}
\end{align}
Notice that at the moment we have not specified the angle-dependent cut $\CB$ on $\CH^+$, so there is currently no angle-dependence in the above formula. To restrict ourselves to the induced radius on the cut, we use \eqref{R0-map} to write the induced radial coordinate in terms of Gaussian null time $\tu$, so that
\begin{align}\label{r-plus}
\begin{split}
	& \p_u a_{\CH^+}(u,z,\bz)\big|_{u = L_0 - 2\Phi(U,z,\bz)} =  - \g_{z\bz} \Phi(\tu,z,\bz) .
\end{split}
\end{align}
As we are only interested in the rate of change of the local area element at the boundary $\CB$, we take the limit
\begin{align}\label{du-a}
\begin{split}
	\p_u a_{\CB}(z,\bz) \equiv \lim_{\tu \to -\infty} \p_u a_{\CH^+} (u,z,\bz)\big|_{u = L_0 - 2\Phi(U,z,\bz)} = - \g_{z\bz} L(z,\bz),
\end{split}
\end{align}
where in the last equality we used \eqref{bdy-cond} with $(\tz,\tbz)$ replaced with $(z,\bz)$. Now, if we parametrize the boundary $\CB$ of an undeformed cut by $\bar L(z,\bz)$, the rate of change of the local area element is just \eqref{du-a} with $L(z,\bz)$ replaced by $\bar L(z,\bz)$. It follows the rate of change of the local area element \emph{fluctuation} at the boundary $\CB$ is simply
\begin{align}\label{CD-fin1}
\begin{split}
	\p_u \D a_\CB(z,\bz) \equiv - \g_{z\bz}( L(z,\bz) - \bar L(z,\bz)) = - \g_{z\bz} \e(z,\bz),
\end{split}
\end{align}
where we used the definition of $\e(z,\bz)$ given in \eqref{eps-def}.\footnote{We remark that this is \emph{not} the same type of area fluctuations considered in \cite{Ciambelli:2025fbo}. In that case, the area fluctuations of the bifurcate horizon $\CB$ were chosen so that the causal diamond itself changed. In this paper, the area fluctuations are those of the transverse area of the cut $\CB$ along a \emph{fixed} null horizon. It is for this reason that if we restrict to the case where $\e(z,\bz)$ is angle-independent, \eqref{CD-fin1} differs from the results in \cite{Ciambelli:2025fbo} by a factor of 2. \label{fn:length-fluc}}

Next, we compute the fluctuation of the celestial sphere located at $\SI^+_-$ due to soft gravitons for the case of asymptotically flat gravity. We can solve \eqref{a-def} with a nontrivial $m_B$ order by order in the large-$r$ limit, and the result is 
\begin{align}
\begin{split}
	r_\AFG(u,z,\bz)^2 = \frac{1}{4}(u_0-u)^2 - 2\int_u^\infty \dt u\, m_B(u,z,\bz) + \cdots.
\end{split}
\end{align}
We can then write the local area element to be
\begin{align}\label{r-soln}
\begin{split}
	a_{\AFG}(u,z,\bz) &\equiv \g_{z\bz} r_{\AFG}(u,z,\bz)^2 \\
	&= \frac{1}{4}\g_{z\bz} (u_0-u)^2 - 2 \g_{z\bz} \int_{u}^\infty \dt u\, m_B(u,z,\bz) + \cdots ,
\end{split}
\end{align}
where $\cdots$ indicates subleading terms in the large-$r$ limit. Now, the term $\frac{1}{4}\g_{z\bz}(u_0-u)^2$ is a divergent reference local area element corresponding to a flat spacetime with $m_B = 0$.\footnote{One can set $u_0 = L_0$ and then take $L_0 \to \infty$.} It is then clear from \eqref{r-soln} that the local area variation from the reference local area is just the integral involving $m_B$, so that
\begin{align}
\begin{split}
	\D a_\AFG(u,z,\bz) &\equiv a_\AFG(u,z,\bz) - \frac{1}{4}\g_{z\bz} (u-u_0)^2 \\
	&= -2\g_{z\bz}\int_{u}^\infty \dt u\,m_B(u,z,\bz).
\end{split}
\end{align}
The rate at which the local area changes at the boundary $\CB$ is then simply
\begin{align}\label{AFG-fin1}
\begin{split}
	\p_u \D a_{\AFG}(u,z,\bz) \big|_{u \to -\infty} = 2 \g_{z\bz} m_B(-\infty,z,\bz).
\end{split}
\end{align}
Matching the rate of change of the local area element using \eqref{CD-fin1} and \eqref{AFG-fin1}, we can construct the map $\Psi$ such that 
\begin{align}\label{mB-eps}
\begin{split}
	\Psi\big[ m_B(-\infty,z,\bz) \big] = - \frac{1}{2} \e(z,\bz)  .
\end{split}
\end{align}
Since $\Psi$ is a symplectomorphism, it preserves the symplectic bracket. This will allow us to later fix the relationship between the Goldstone mode $C$ and the null time offset $\a$.

However, let us first rewrite \eqref{mB-eps} in terms of the soft mode $N(z,\bz)$. Substituting the shear profile \eqref{soft-shear} into \eqref{eq:mBconstraint}, we get
\begin{align}\label{Du-bar0}
\begin{split}
	\p_u m_B(u,z,\bz) &= \frac{1}{2} (\g^{z\bz})^2D_z^2 D_\bz^2 N(z,\bz) \d(u-u_s) + \cdots ,
\end{split}
\end{align}
where $\cdots$ indicates $\CO(N^2)$ terms.\footnote{The $\CO(N^2)$ terms capture the hard contribution to gravitational radiation, which can be neglected if we are only interested in studying the soft modes.} Substituting \eqref{2-deriv} into \eqref{Du-bar0}, we get 
\begin{align}\label{Du-bar}
\begin{split}
	\p_u m_B(u,z,\bz) = \frac{1}{8} \Box (\Box + 2) N(z,\bz) \d(u-u_s) + \cdots .
\end{split}
\end{align}
Integrating \eqref{Du-bar0} over $\SI^+$ and with the boundary condition $m_B|_{\SI^+_+} = 0$, which is also the boundary condition assumed in \cite{Kapec:2016aqd}, we get
\begin{align}
\begin{split}
	m_B(u,z,\bz) = - \frac{1}{8}\Box(\Box+2) N(z,\bz) \big( 1 - \th(u-u_s) \big) .
\end{split}
\end{align}
Evaluating this at $\SI^+_-$ by setting $u = -\infty$, we get
\begin{align}\label{mB-N}
\begin{split}
	m_B(-\infty,z,\bz) = - \frac{1}{8} \Box ( \Box+2) N(z,\bz).
\end{split}
\end{align}
Substituting \eqref{mB-N} into \eqref{mB-eps}, we arrive at linear order in $N$
\begin{align}\label{eps-N}
\begin{split}
	\Psi\big[ \Box ( \Box+2) N(z,\bz) \big] = 4\e(z,\bz) . 
\end{split}
\end{align}

Next, we turn to using the symplectic analysis to relate $C(z,\bz)$ to the mode $\a(z,\bz)$. Since we want $\Psi$ to preserve the bracket, we require
\begin{align}\label{C-L}
\begin{split}
	&  \Big\{ \Psi\big[ C(z,\bz) \big], \Psi\big[\Box (\Box + 2) N(w,\bw) \big]  \Big\} = \{  C(z,\bz), \Box (\Box + 2) N(w,\bw) \} \\
	\implies\quad &   \Big\{ \Psi\big[ C(z,\bz) \big], \e(w,\bw) \Big\} = - 8\pi G \g^{z\bz} \d^2(z-w) \\
	\implies\quad &  \Psi\big[ C(z,\bz) \big] = 2\a(z,\bz) L(z,\bz) + h\big[L(z,\bz)\big] , 
\end{split}
\end{align}
where in the second line we substituted in \eqref{soft-comm} and \eqref{eps-N}, and in the last line we used \eqref{final-bracket} as well as noted the fact we can add an arbitrary phase space functional $h[L(z,\bz)]$ as the bracket between $L(z,\bz)$ and $\e(w,\bw)$ vanishes. The choice of $h$ in \eqref{C-L} corresponds to choosing a particular identification between the vacuum labeled by $C$ and that labeled by $\a$, both of which are not physically observable. Therefore, we can choose the functional $h$ such that the $\a =0$ state corresponds to the $C=0$ state. This involves setting $h[L(z,\bz)] = 0$, so that our final identification is
\begin{align}\label{CL-fin}
\begin{split}
	 \Psi\big[ C(z,\bz)\big] = 2 \a(z,\bz) L(z,\bz). 
\end{split}
\end{align}
This completes our identification between the phase space of cuts on a finite null hypersurface in a Minkowski background and that of asymptotically flat spacetime in four spacetime dimensions.

\section{Discussion and future directions} \label{sec:discussion}

Our goal in this paper is to construct a natural isomorphism between the phase space of the leading infrared sector of an asymptotically flat spacetime and that of cuts on a finite null hypersurface in a Minkowski background. We achieved this by first matching the rate of change of area fluctuations in both cases. This allowed us to identify the leading soft graviton mode $N(z,\bz)$ with the length fluctuation mode $\e(z,\bz)$ of a cut on the finite null hypersurface. We then used the symplectic structure to determine the map between the supertranslation Goldstone $C(z,\bz)$ and the null time offset mode $\a(z,\bz)$. We view this paper as the generalization of the analysis performed in \cite{Bub:2024nan, Ciambelli:2025fbo} to angle-dependent length fluctuations, although the type of spherically symmetric length fluctuations studied in \cite{Bub:2024nan, Ciambelli:2025fbo} were slightly different (see Footnote~\ref{fn:length-fluc}).

Physically, asymptotically flat spacetimes differ quite dramatically from the finite subregions of Minkowski spacetime considered in Section~\ref{subsec:mink}. This can understood rather directly by recognizing that the metrics \eqref{eq:GN} and \eqref{eq:bondi} are \emph{not} diffeomorphic. Quite remarkably however, as we demonstrated above, the soft sector of asymptotically flat gravity is closely related to the spin-0 sector of the finite Minkowski region from the \emph{phase space} point of view. In some sense, this observation underscores a certain universality pertaining to the vacuum sector in gravity \cite{Klinger:2025tvg}, its relation to dynamical vacuum transitions, and related information theoretic considerations \cite{Prabhu:2022zcr,AliAhmad:2024saq,AliAhmad:2025oli,Danielson:2025aji}.

Indeed, as was explored in \cite{Verlinde:2022hhs, Bak:2024kzk, Zhang:2023mkf}, the fluctuations of the size of the transverse cut of a finite null hypersurface in Minkowski background can be modeled using null shockwaves. These shockwaves do not need to be sourced by a matter stress tensor, but can arise from the fluctuations of the Bondi mass aspect itself \cite{He:2023qha, He:2024vlp}. As a result, there is in fact a relationship between the soft graviton mode and the momentum of such a shockwave, which was determined in \cite{He:2023qha, He:2024vlp} to be
\begin{align}
\begin{split}
	P_u(z,\bz) = \frac{1}{32\pi G} \Box( \Box+2) N(z,\bz).
\end{split}
\end{align}
Using \eqref{eps-N}, it then follows the precise relationship between the length fluctuation of the cut $\CB$ and the shockwave momentum is\footnote{As was emphasized in Footnote~\ref{fn:length-fluc}, the length fluctuations considered here differs from those considered in \cite{Ciambelli:2025fbo}, which is why there is a factor of 2 difference in \eqref{P-eps} when compared with the results of \cite{Ciambelli:2025fbo}. } 
\begin{align}\label{P-eps}
	\Psi\big[ P_u(z,\bz) \big] = \frac{1}{8\pi G} \e(z,\bz) .
\end{align}
It was further argued in \cite{He:2023qha, He:2024vlp} that there is also a natural identification between the Goldstone mode and the shockwave null shift, which is
\begin{align}
	X^u(z,\bz) &= - C(z,\bz).
\end{align}	
Using \eqref{CL-fin}, we see that we can identify the null time shift $\a(z,\bz)$ of the cut $\CB$ with the shockwave null shift via
\begin{align}\label{X-alpha}
\begin{split}
	\Psi\big[ X^u(z,\bz) \big] = - 2\a(z,\bz)L(z,\bz) .
\end{split}
\end{align}
Of course, the ambiguity involving the functional $h[L]$ in \eqref{C-L} still persists here, so we should really view \eqref{X-alpha} as a choice, fixed by having the shockwave with null shift $X^u=0$ corresponding to the initial condition $\a=0$. Importantly, though, because only changes in $X^u$ are observable (or equivalently, changes in $\a$), shifting the identification given in \eqref{X-alpha} by $h[L]$ does not affect any observables.

There are a couple natural future directions that need to be explored. From a formal perspective, we would like to canonically quantize the analysis performed in this paper. All of our results above are classical statements, and we would like to replace the symplectic brackets with quantum commutators. However, because the identification \eqref{CL-fin} is nonlinear, there are operator ordering ambiguities between $\a$ and $L$. A natural choice would be the Weyl ordering, so that the generalization of \eqref{CL-fin} to quantum operators would be the symmetrized sum
\begin{align}
	\Psi \big[ \hat C(z,\bz) \big] = \hat \a(z,\bz) \hat L(z,\bz) + \hat L(z,\bz) \hat \a(z,\bz).
\end{align}
This will ensure that the operator identified with the Goldstone $C$ is now Hermitian. It will be interesting to explore the consequences of such an identification.

Ultimately, a major motivation for this work is to tie our analysis to interferometer observables. To this end, relaxing spherical symmetry is an important step forward, but our current analysis remains largely formal. Nevertheless, we are hopeful that the correspondence identified in this paper can be used to leverage results from asymptotically flat spacetimes to make concrete predictions for experimental setups probing length fluctuations in finite spacetime regions.

\section*{Acknowledgements}

We would like to thank Sang-Eon Bak, Cynthia Keeler, and Prahar Mitra for useful conversations. Research at Perimeter Institute is supported in part by the Government of Canada through the Department of Innovation, Science and Economic Development Canada and by the Province of Ontario through the Ministry of Colleges and Universities. L.C. is supported by the Celestial Holography Simons collaboration.  T.H., M.S.K., and K.Z. are supported by the Heising-Simons Foundation “Observational Signatures of Quantum Gravity” collaboration grant 2021-2817, the U.S. Department of Energy, Office of Science, Office of High Energy Physics, under Award No. DE-SC0011632, and the Walter Burke Institute for Theoretical Physics. K.Z. is also supported by a Simons Investigator award.

\appendix

\section{Brief review of Carrollian structure}\label{app:carroll-review}

In this appendix, we provide the reader a brief review of the Carrollian formalism, which is particularly useful for studying null hypersurfaces and provides an intrinsic geometric classification that also naturally connects to the ambient spacetime picture (see \cite{Ciambelli:2025unn} for more details). In particular, we will see how our Gaussian null metric \eqref{eq:GN} emerges from the more general Carrollian metric.

Given a $(d+1)$-dimensional manifold $\CH^s$, oftentimes called a stretched horizon, let the intrinsic coordinates on it be denoted $x^a = (\tu, x^i)$. A (ruled) stretched Carrollian structure on it is given by $(h_{ab}, \ell^a, k_a, \tr_0)$, where $h_{ab}$ is the metric, and $\ell^a$ a vector field, $k_a$ a one-form, and $R_0$ a scalar. These quantities comprises the data on $\CH^s$ and obeys
\begin{align}\label{eq:ell-k}
	\ell^a k_a = 1 , \qquad h_{ab} \ell^a = -2 \tr_0 k_b .
\end{align}
Notice that this implies that $\ell^a$ cannot vanish anywhere, and indeed we have
\begin{align}\label{ell-norm}
	h_{ab}\ell^a \ell^b = - 2\tr_0.
\end{align}
We may view $\ell^a$ as the tangent vector on $\CH^s$, and so \eqref{ell-norm} implies that $\tr_0$ controls the causal nature of the horizon and is sometimes referred to as the stretching parameter. In the case $\tr_0 = 0$, the metric $h_{ab}$ becomes the degenerate Carrollian metric $q_{ab}$, and $\CH^s$ becomes a null horizon $\CH^{+}$. In this case, the ruled Carrollian structure on the null manifold $\CH^{+}$ is given by the data $(q_{ab}, \ell^a, k_a)$.

We are now interested in embedding the stretched horizon $\CH^s$ into an $(d+2)$-dimensional ambient spacetime $\CM$. We can realize a foliation of the spacetime via a one-parameter family of maps
\begin{align}\label{eq:sCarrEmb}
	j_{\tr_0} : \CH^s \to \CM,
\end{align}
where $j_{\tr_0}$ is the embedding map for the stretched horizon $\CH^s$ given above. These maps may be chosen such that $j_0(\CH^+) \subset \CM$ is null while $j_{\tr_0}(\CH^s)$ is timelike for $\tr_0 > 0$. It is then convenient to introduce an adapted bulk coordinate $\tr$ such that $j_{\tr_0}(\CH^s) \subset \CM$ are the level sets $\tr=\tr_0$. It follows the family of embedding maps given in \eqref{eq:sCarrEmb} can be characterized by the normal one-form $n \equiv \dt \tr$ and the tangent bundle dual vector field $k \equiv \p_\tr$.\footnote{Note that by \eqref{eq:ell-k},  the pullback of the metric dual one-form $k$ to $\CH^s$ is dual to the tangent vector $\ell$.} Along with the spacetime metric $g_{\mu\nu}$, the collection $(\CM, g_{\mu\nu}, k^\mu, n_\nu)$ provides a bulk rigging structure and allows us to canonically project to the stretched horizon $\CH^s$ \cite{Mars:1993mj}. From this point of view, an ambient metric that pulls back to a stretched Carrollian manifold can be written in the form \cite{Freidel:2022vjq}
\begin{align}\label{eq:sCarrMetric}
\begin{split}
	\dt s^2 &= 2 e^\a ( \dt \tu - \b_i e^i )\big(\dt \tr - \rho e^\a (\dt \tu - \b_j e^j ) \big) + q_{ij} e^i e^j ,
\end{split}
\end{align}
where $\rho$ is a general function, and $e^i \equiv J^i{}_j (\dt \tx^j - V^j \dt \tu )$ is a co-frame for the $d$-dimensional space orthogonal to the foiliation, i.e., the cuts of the Carrollian manifold. The above metric is parametrized by the scale function $\a$, the Carrollian connection $\b_i$, the velocity vector field $V^i$, and the $d$-dimensional metric $q_{ij}$, which are all a priori generic functions in the bulk. We can pull back the general Carrollian metric to the following induced line element on a constant $\tr = \tr_0$ leaf, obtaining
\begin{align}
\begin{split}
	\dt s^2\big|_{\tr= \tr_0} = -2 \tr_0 e^{2\a}( \dt \tu - \b_i e^i ) ( \dt \tu - \b_j e^j) + q_{ij} e^i e^j.
\end{split}
\end{align}
Thus, we see that for $\tr_0 > 0$, this is a timelike hypersurface, whereas for  $\tr_0 = 0$, this is a null hypersurface. 

Comparing \eqref{eq:sCarrMetric} with \eqref{eq:GN}, we see that the Gaussian null metric is the subclass of Carrollian metrics where to linear order in $\tr$
\begin{align}
	\a = 0, \qquad \rho(\tr) = \ka \tr, \qquad \b_i = 0 , \qquad J^i{}_{j} = \d^i_j , \qquad V_i = 2\tr \pi_i .
\end{align}
This arises due to us specializing to the case where only the spin-0 and spin-1 degrees are treated as dynamical, and is the starting point for our analysis in the main text.

\section{Solving Raychaudhuri and Damour equations without shear} \label{app:RayDam}

In this appendix, we provide the solutions to the Raychaudhuri and Damour equations with vanishing shear $\s_{ij} = 0$. These two equations are given in \eqref{eq:ray} and \eqref{eq:dam}, and we rewrite them here for convenience as
\begin{align}
	\p_\tu \th  &= \ka \th - \frac{1}{d} \th^2 \label{gen-ray2} \\
	\p_\tu \pi_i + \th\pi_i &= \p_i\ka + \frac{d-1}{d}\p_i\th. \label{gen-dam2}
\end{align}
We are interested in the case when $\ka \not=0$. Furthermore, as we argued in Section~\ref{subsec:mink}, we can perform a small gauge-fixing so that $\ka(\tu,\vec\tx) \equiv \ka(\vec\tx)$, i.e., it is time-independent (see Footnote~\ref{fn:small-gauge}). With these assumptions, we can solve \eqref{gen-ray2} for $\th$ and obtain
\begin{align}\label{th-soln}
	\th(\tu,\vec\tx) = 
		\frac{\ka(\vec\tx) d }{ 1 - e^{-\ka(\vec\tx)(\tu - \tu_0) - \a(\vec\tx)}} ,
\end{align}
where $\a(\vec\tx)$ is an integration constant defined via
\begin{align}\label{alpha-def}
\begin{split}
	\th(\tu_0 , \vec\tx) = \frac{\ka(\vec\tx) d }{1 - e^{- \a(\vec\tx) }} \quad\implies\quad e^{-\a(\vec\tx)} = 1 - \frac{\ka(\vec\tx) d}{\th(\tu_0,\vec\tx)}  ,
\end{split}
\end{align}
and $\tu_0$ is the initial time (possibly $-\infty$) so that $U > \tu_0$. Recalling the definition of $\th$ given in \eqref{eq:theta}, \eqref{th-soln} becomes the differential equation
\begin{align}
\begin{split}
	\p_\tu\log\varphi(\tu,\vec\tx) = \frac{\ka(\vec\tx) d }{ 1 - e^{-\ka(\vec\tx)(\tu-\tu_0) - \a(\vec\tx)}}   ,
\end{split}
\end{align}
which can be straightforwardly solved to be
\begin{align}\label{Omega-soln}
\begin{split}
	\varphi(\tu,\vec\tx) = \varphi(\tu_0,\vec\tx) \Bigg( \frac{e^{\ka(\vec\tx) (\tu - \tu_0) + \a(\vec\tx)} - 1}{ e^{\a(\vec\tx)} - 1 } \Bigg)^d  . 
\end{split}
\end{align}
Notice that because $\ka > 0$ and $\tu > \tu_0$ by construction, the above equation is always nonnegative. This fully fixes the form of $\varphi$ in terms of the initial conditions.

Next, let us use the above results to solve the Damour equation \eqref{gen-dam2}. Substituting \eqref{th-soln} into \eqref{gen-dam2} and integrating over $U$, we get 
\begin{align}\label{pi-soln}
\begin{split}
	\pi_i(\tu,\vec\tx) &= \frac{C_i(\vec\tx)}{ \big( e^{\ka(\tu-\tu_0) + \a(\vec\tx)} - 1 \big)^{d}} +  \p_i\log\ka(\vec\tx) \\
	&\qquad - e^{-\ka(\vec\tx)(\tu-\tu_0) - \a(\vec\tx)} \p_i \log \Big|  e^{\ka(\vec\tx)(\tu-\tu_0) + \a(\vec\tx)} - 1 \Big| ,
\end{split}
\end{align}
where $C_i(\vec\tx)$ is an integration constant. To fix $C_i(\vec\tx)$, note that at $\tu=\tu_0$, we have
\begin{align}\label{C-def}
\begin{split}
	&\pi_i(\tu_0,\vec\tx) = \frac{C_i(\vec\tx)}{\big(  e^{\a(\vec\tx)} - 1 \big)^{d} } + \p_i\log \ka(\vec\tx) - e^{-\a(\vec\tx)}\p_i\log\big|  e^{\a(\vec\tx)} - 1 \big| \\
	\implies\quad & C_i(\vec\tx) = \big( e^{\a(\vec\tx)} - 1 \big)^d \Big[ \pi_i(\tu_0,\vec\tx) - \p_i \log\ka(\vec\tx) + e^{-\a(\vec\tx)}\p_i \log \big|  e^{\a(\vec\tx)}  - 1 \big| \Big] .
\end{split}
\end{align}
It is clear from \eqref{pi-soln} that $\pi_i$ is completely determined by the initial conditions and transverse derivatives of the data $\ka(\vec\tx)$. In fact, it is also instructive to solve for $\pi_i$ in \eqref{gen-dam2} directly without using immediately \eqref{th-soln}, and the result is
\begin{align}\label{pi-soln2}
\begin{split}
	\pi_i(\tu,\vec\tx) &= \frac{1}{\varphi(\tu,\vec\tx)} \bigg[ \pi_i(\tu_0,\vec\tx)\varphi(\tu_0,\vec\tx) + \int_{\tu_0}^\tu \dt U'\, \varphi(U',\vec\tx) \p_i \mu(\tu,\vec\tx) \bigg] 
\end{split}
\end{align}
where we used the definition of $\th$ given in \eqref{eq:theta}, and the definition of $\mu$ given in \eqref{eq:mu}. 

It is clear from \eqref{pi-soln2} that even though $\pi_i$ is determined wholly in terms of transverse derivatives of the boundary data $\ka(\vec\tx)$ and $\varphi(\tu,\vec\tx)$, it is not a \emph{total} transverse derivative in general. However, if the time interval $[\tu_0,\tu]$ is small enough such that $\varphi$ is roughly constant in this interval, then we can replace $\varphi(\tu,\vec\tx)$ with $\varphi(\tu_0,\vec\tx)$, so that the above equation simplifies to
\begin{align}
\begin{split}
	\pi_i(\tu,\vec\tx) &= \pi_i(\tu_0,\vec\tx) + \p_i \int_{\tu_0}^{\tu} \dt \tu' \,\mu(\tu',\vec\tx) \\
	&= \pi_i(\tu_0,\vec\tx) + \p_i \int_{\tu_0}^{\tu} \dt \tu' \,\p_{\tu'} \Big( \log\varphi(\tu',\vec\tx) + \log|\th(\tu',\vec\tx)| \Big) ,
\end{split}
\end{align}
where in the second line we substituted in \eqref{eq:theta} and \eqref{mu-onshell}. This implies in this small time interval we have
\begin{align}\label{pi-lin}
	\pi_i(\tu,\vec\tx) = \p_i \big( \log\varphi(\tu,\vec\tx) + \log|\th(\tu,\vec\tx)| \big) ,
\end{align}
so that $\pi_i$ is indeed a total transverse derivative.\footnote{Note that since $\Phi$ is very slowly varying in the interval, $\th$ is very small, so $\log|\th|$ is very large and dominates in \eqref{pi-lin}.} Perhaps what is surprising is that for the case of interest for us considered in Section~\ref{subsec:mink}, \eqref{pi-soln2} reduces to a total transverse derivative for the particular choice of $\ka$ without having to restrict ourselves to the above small interval in time. To see this, we set $d=2$ and substitute in $\varphi$ and $\ka$ from \eqref{eq:data} and $\th$ from \eqref{th-Phi}, so that \eqref{pi-soln2} becomes
\begin{align}
\begin{split}
	\pi_\tz(\tu,\tz,\tbz) &= \frac{1}{\varphi(\tu,\tz,\tbz)} \Bigg[ \pi_\tz(\tu_0,\tz,\tbz)\varphi(\tu_0,\tz,\tbz) + \int_{\tu_0}^\tu \dt U'\, \varphi(U',\tz,\tbz) \p_\tz \mu(\tu',\tz,\tbz) \Bigg] \\
	 &= \frac{\Phi(\tu_0,\tz,\tbz)^2}{\Phi(\tu,\tz,\tbz)^2}  \bigg[ \pi_\tz(\tu_0,\tz,\tbz) + \frac{\p_\tz \Phi(\tu_0,\tz,\tbz)}{\Phi(\tu_0,\tz,\tbz)}  -  \frac{\p_{\tu_0}\p_\tz\Phi(\tu_0,\tz,\tbz)}{\p_{\tu_0}\Phi(\tu_0,\tz,\tbz)} \bigg] \\
	 &\qquad - \frac{\p_\tz\Phi(\tu,\tz,\tbz) }{\Phi(\tu,\tz,\tbz)}  + \frac{\p_{\tu}\p_\tz\Phi(\tu,\tz,\tbz)}{\p_{\tu}\Phi(\tu,\tz,\tbz)} \\
	 &= \frac{\Phi(\tu_0,\tz,\tbz)^2}{\Phi(\tu,\tz,\tbz)^2}  \bigg[ \pi_\tz(\tu_0,\tz,\tbz) - \p_\tz \log\big| \p_{\tu_0} \log\Phi(\tu_0,\tz,\tbz) \big| \bigg] \\
	 &\qquad + \p_\tz\log\big|\p_\tu\log\Phi(\tu,\tz,\tbz)\big| .
\end{split}
\end{align}
Thus we immediately see this implies that
\begin{align}
	\pi_\tz = \p_\tz\log\big|\p_\tu\log\Phi(\tu,\tz,\tbz)\big| ,
\end{align}
which is indeed a total transverse derivative and is consistent with \eqref{eq:data}.

\section{Deriving the diffeomorphism}\label{app:diffeo}

In this appendix, we will derive the diffeomorphism used in Section~\ref{subsec:mink} that allows us to describe a non-spherically symmetric cut of a finite null hypersurface in four-dimensional Minkowski spacetime using Gaussian null coordinates. We begin by recalling that the Minkowski metric written in retarded coordinates is given by
\begin{align}\label{eq:mink-app}
	\dt s^2 = - \dt u^2 - 2\dt u\,\dt r + r^2 \g_{z\bz} \dt x^A \dt x^B,
\end{align}
where $u = t - r$ is the retarded null time, $x^A$ the Minkowski angular coordinates $(z, \bz)$, and 
\begin{align}\label{round-met}
	\g_{AB} = \begin{pmatrix}
		0 &  \frac{2}{(1+z \bz)^2} \\
		 \frac{2}{(1+z \bz)^2} & 0
	\end{pmatrix} 
\end{align}
is the round metric on the unit sphere. It is useful for later convenience to denote the inverse round metric $\g^{AB} \equiv (\g_{AB})^{-1}$. We now construct a diffeomorphism that takes us to Gaussian null coordinates $(\tu,\tr,\vec\tx)$, where $\tx^i$ labels the Gaussian null angular coordinates $(\tz,\tbz)$. 

As the symplectic form lives on the bifurcate horizon $\CB$ of the causal diamond, it suffices to work out the diffeomorphism in the neighborhood of the future horizon $\CH^+$, given by $\tr=0$ in Gaussian null coordinates. We can then define the following coordinate map between the two metrics order by order in a small $\tr$ expansion:
\begin{align}\label{diffeo}
\begin{split}
	u(\tu,\tr,\tz,\tbz) &= \Xi(\tu,\tz,\tbz) + u^\1(\tu,\tz,\tbz)\tr + u^\2(\tu,\tz,\tbz)\tr^2 + \CO(\tr^3) \\
	r(\tu,\tr,\tz,\tbz) &= \Phi(\tu,\tz,\tbz) + r^\1(\tu,\tz,\tbz)\tr +  r^\2(\tu,\tz,\tbz)\tr^2 + \CO(\tr^3) \\
	z(\tu,\tr,\tz,\tbz) &= \tz + z^\1(\tu,\tz,\tbz)\tr + z^\2(\tu,\tz,\tbz)\tr^2 + \CO(\tr^3) \\
	\g_{z\bz}(\tu,\tr,\tz,\tbz) &= \g_{\tz\tbz} + \g_{z\bz}^\1(\tu,\tz,\tbz) \tr + \CO(\tr^2) .
\end{split}
\end{align}
We are interested in keeping track of $\CO(\tr)$ terms in the Gaussian null metric, so we need to keep track of $\CO(\tr^2)$ terms in $u,r,z$ above since the exterior derivative acting on $\tr^2$ removes a factor of $\tr$. However, it suffices to keep $\g_{z\bz}$ to $\CO(\tr)$ since no exterior derivatives act on $\g_{z\bz}$. Indeed, substituting \eqref{diffeo} into \eqref{round-met}, we can immediately determine $\g_{z\bz}^\1$ in terms of the functions $z^\1$ and $\bz^\1$:
\begin{align}
\begin{split}
	\g_{z\bz}(\tu,\tr,\tz,\tbz) &= \frac{2}{(1+z\bz)^2} = \g_{\tz\tbz} + \g_{\tz\tbz} \big( \G^{\tbz}_{\tbz\tbz} \bz^\1 + \G^{\tz}_{\tz\tz} z^\1 \big) \tr + \CO(\tr^2),
\end{split}
\end{align}
where $\G^{\tz}_{\tz\tz} = - \frac{2\tbz}{1+\tz\tbz}$ is the Christoffel symbol associated to the round metric. The fact that $\g_{z\bz}$ and $\g_{\tz\tbz}$ agree to leading order in a small $\tr$ expansion illustrates that to leading order we are not transforming the angular coordinates, and in particular on the horizon $\CH^+$ where $\tr=0$ they remain the same under the coordinate map. 

In Gaussian null coordinates, the line element in four dimensions has the form (see \eqref{eq:GN})
\begin{align}\label{GN-metric}
\begin{split}
	\dt s^2 &= - 2\ka \tr\,\dt \tu^2 + 2 \, \dt \tu\,\dt \tr - 4 \tr \pi_i\,\dt \tu\,\dt \tx^i  + q_{ij} \,\dt \tx^i\,\dt \tx^j + \CO(\tr^2) \\
	&= - 2\ka \tr\,\dt \tu^2 + 2 \, \dt \tu\,\dt \tr - 4 \tr \pi_i\,\dt \tu\,\dt \tx^i  +  \Phi^2 \big( \g_{ij} + \tr \g_{ij}^\1  \big) \,\dt \tx^i\,\dt \tx^j + \CO(\tr^2).
\end{split}
\end{align}
where 
\begin{align}\label{qij-exp}
	q_{ij}(\tu,\tr,\tz,\tbz) = \Phi^2(\tu,\tz,\tbz)\big( \g_{ij} + R \g_{ij}^\1(\tu,\tz,\tbz)  \big) + \CO(\tr^2).
\end{align}
Note that \eqref{GN-metric} is precisely \eqref{eq:GN} with the identification $\varphi = \Phi^2$. We now want to determine the form of the data in \eqref{GN-metric} (such as $\ka, \pi_i, \Phi$, etc.) so that the metric \eqref{GN-metric} is diffeomorphic to \eqref{eq:mink-app}. This simply involves substituting the coordinate map \eqref{diffeo} into \eqref{eq:mink-app} and matching it to the Gaussian null metric \eqref{GN-metric}. We will do this order by order in $\tr$. For notational simplicity, we will henceforth adopt the notation where all functions are implicitly a function of $(\tu,\tz,\tbz)$ unless otherwise specified.  

First, we require that $g_{\tu\tu} = 0$ at $\CO(\tr^0)$. This implies
\begin{align}\label{U0}
\begin{split}
	g_{\tu\tu}\big|_{\CO(\tr^0)} = 0 \quad\implies\quad \Xi = -2\Phi + L_0(\tz,\tbz) ,
\end{split}
\end{align}
where $L_0(\tz,\tbz)$ is an arbitrary function of the angular coordinates. Next, we require that $g_{\tr\tr} = 0$ to all orders in $\tr$. Requiring this at $\CO(\tr^0)$ implies
\begin{align}\label{R1}
\begin{split}
	g_{\tr\tr}\big|_{\CO(1)} = 0 \quad\implies\quad r^\1 = - \frac{1}{2} u^\1 + \frac{\bz^\1 z^\1 \Phi^2 \g_{\tz\tbz}}{u^\1}.
\end{split}
\end{align}
Further requiring $g_{\tr\tr} = 0$ at $\CO(\tr)$, we get
\begin{align}\label{R2}
\begin{split}
	g_{\tr\tr}\big|_{\CO(\tr)} = 0 \quad\implies\quad r^\2 &= \frac{1}{2}\Bigg[ \frac{\Phi^2}{u^\1} \Big( 2z^\1\bz^\2 \g_{\tz\tbz} +  2 z^\2 \bz^\1 \g_{\tz\tbz} + z^\1 \bz^\1 \g^\1_{z\bz} \Big) \\
	&\qquad - u^\2 \Bigg( 1 + \frac{2 z^\1 \bz^\1 \g_{\tz\tbz} \Phi^2}{(u^\1)^2}  \Bigg)   \Bigg]  .
\end{split}
\end{align}
Next, we require $g_{\tu\tr} = 2$ to all orders in $\tr$. First imposing this at $\CO(\tr^0)$, we get
\begin{align}\label{U1}
\begin{split}
	g_{\tu\tr}\big|_{\CO(\tr^0)} = 2 \quad\implies\quad u^\1 = 2 \g_{\tz\tbz} z^\1 \bz^\1 \Phi^2 \p_\tu \Phi.
\end{split}
\end{align}
We will return to imposing $g_{\tu\tr} = 2$ to $\CO(\tr)$, as it is more convenient to next require $g_{\tu\tz} = \CO(\tr)$. This constraint results in
\begin{align}\label{Lambda}
\begin{split}
	g_{\tu\tz}\big|_{\CO(\tr^0)} = 0 \quad\implies\quad L_0(\tz,\tbz) = L_0,
\end{split}
\end{align}
where $L_0$ is now just a constant.

Let us continue with imposing $g_{\tr\tz} = 0$, which holds to all orders in $\tr$. Requiring this at $\CO(\tr^0)$ implies
\begin{align}\label{Z1}
\begin{split}
	g_{\tr\tz}\big|_{\CO(\tr^0)} = 0 \quad\implies\quad \bz^\1 = -\frac{\p_\tz \Phi}{\g_{\tz\tbz} \Phi^2 \p_\tu \Phi},
\end{split}
\end{align}
and similarly $z^\1$ is given by the complex conjugate of \eqref{Z1}. Further requiring $g_{\tr\tz}=0$ at $\CO(\tr)$, we get
\begin{align}\label{U2}
\begin{split}
	g_{\tr\tz} \big|_{\CO(\tr)} = 0 \quad\implies\quad u^\2 &= - \p_\tz \Phi \Bigg[ 2 z^\2 + \frac{( \p_\tbz \Phi)^2 }{\g_{\tz\tbz}^3 \Phi^5 ( \p_\tu \Phi)^2 } \big( \Phi \p_\tz\g_{\tz\tbz} + 2 \g_{\tz\tbz} \p_\tz \Phi \big)  \Bigg]  .
\end{split}
\end{align}
Finally, we return to requiring $g_{\tu\tr} = 2$ to $\CO(\tr)$, i.e., the $\CO(\tr)$ part of $g_{\tu\tr}$ vanishes. This yields
\begin{align}\label{Z2}
\begin{split}
	g_{\tu\tr} \big|_{\CO(\tr)} = 0 \quad\implies\quad \bz^\2 &= \frac{\p_\tz \Phi}{2\g_{\tz\tbz}^3 \Phi^4 (\p_\tu \Phi)^2} \Big( \p_\tz\g_{\tz\tbz}\p_\tbz \Phi + \g_{\tz\tbz} \g_{z\bz}^\1 \Phi^2\p_\tu\Phi   \Big) ,
\end{split}
\end{align}
and similarly $z^\2$ is given by the complex conjugate of \eqref{Z2}.

We are finally able to derive the form of the Gaussian null line element \eqref{GN-metric} that is diffeomorphic to the Minkowski line element \eqref{eq:mink-app}. Substituting the above equalities into \eqref{eq:mink-app}, we arrive at the Gaussian null line element \eqref{GN-metric}, with the following identifications:
\begin{align}\label{asy-data}
\begin{split}
	\varphi &= \Phi^2 \\
	\ka &=  \p_\tu\log |\p_\tu \Phi |  \\
	\pi_\tz &= \p_\tz  \log \big|\p_\tu \log \Phi \big|  \\
	q_{\tz\tz} &= \frac{2\tr}{\Phi\p_\tu\Phi} \big[ 2 (\p_\tz \Phi)^2 - \Phi  D_\tz^2 \Phi  \big] \\
	q_{\tz\tbz} &= \g_{\tz\tbz}\Phi^2 + \frac{2\tr}{\p_\tu\Phi} \bigg( \frac{2\p_\tz\Phi\p_\tbz\Phi}{\Phi} - \p_\tz\p_\tbz\Phi  \bigg) .
\end{split}
\end{align}
Note that the first equality already follows from our discussion below \eqref{qij-exp}. Finally, we write down the full coordinate map for the various metric components in \eqref{diffeo}:
\begin{align}\label{diffeo2}
\begin{split}
	u(\tu,\tr,\tz,\tbz) &= L_0 - 2\Phi + \frac{2\g^{\tz\tbz} \p_\tz\Phi\p_\tbz\Phi}{\Phi^2\p_\tu\Phi}\tr - \frac{2 \big(\g^{\tz\tbz}\p_\tz\Phi \p_\tbz\Phi\big)^2}{\Phi^5(\p_\tu\Phi)^2 }\tr^2 + \cdots \\
	r(\tu,\tr,\tz,\tbz) &=  \Phi + \frac{\Phi^2 - 2\g^{\tz\tbz}\p_\tz\Phi\p_\tbz\Phi}{2\Phi^2\p_\tu\Phi} \tr + \frac{\g^{\tz\tbz} \p_\tz\Phi\p_\tbz \Phi \big( \Phi^2 + 2\g^{\tz\tbz}\p_\tz\Phi\p_\tbz\Phi \big) }{2\Phi^5 (\p_\tu\Phi)^2} \tr^2 + \cdots   \\
	z(\tu,\tr,\tz,\tbz) &=\tz - \frac{\g^{\tz\tbz} \p_\tbz\Phi}{\Phi^2\p_\tu\Phi} \tr - \frac{\G^{\tz}_{\tz\tz}\big(\g^{\tz\tbz}\p_\tbz \Phi\big)^2}{2\Phi^4(\p_\tu\Phi)^2} \tr^2 + \cdots ,
\end{split}
\end{align}
where $\cdots$ indicate subleading $\CO(\tr^3)$ terms, and recall that $L_0$ is an arbitrary spacetime constant.

\bibliography{references_use}{}
\bibliographystyle{utphys}

\end{document}